\documentclass{aa} 
\usepackage{graphicx}
\begin{document}
\def\ffam {\hbox{$\,.\!\!^{\prime}$}}
\def\ffas {\hbox{$\,.\!\!^{\prime\prime}$}}
\def\ffM {\hbox{$\,.\!\!^{\rm M}$}}
\def\ffm {\hbox{$\,.\!\!^{\rm m}$}}
\def\ffs {\hbox{$\,.\!\!^{\rm s}$}}

\title{
Hyperfine Structure in H$^{13}$CO$^{+}$ and $^{13}$CO: measurement, analysis, and
consequences for the study of dark clouds}

\author{Johannes Schmid-Burgk\inst{1}
        \and
        Dirk Muders\inst{1}
        \and
        H. S. P. M\"{u}ller\inst{2}
        \and
        Bethany Brupbacher-Gatehouse\inst{3}
        }

\offprints{J. Schmid-Burgk,
\email{schmid-burgk@mpifr-bonn.mpg.de}}

\institute{
           Max-Planck-Institut f\"ur Radioastronomie, Auf dem H\"ugel
              69, D-53121 Bonn, Germany
           \and
I. Physikalisches Institut, Universit\"{a}t zu K\"{o}ln, D-50937 K\"{o}ln, Germany
           \and
Lab. f\"{u}r Phys. Chemie, ETH-H\"{o}nggerberg, CH-8093 Z\"{u}rich, Switzerland
}

\date{Received date / Accepted date}

\abstract{The magnetic moment of the $^{13}$C nucleus is shown to provide a potentially
useful tool for analysing quiescent cold molecular clouds.
We report discovery of hyperfine structure in the lowest rotational transition of H$^{13}$CO$^{+}$.
The doublet splitting in H$^{13}$CO$^{+}$, observed to be of width 38.5$\pm$5.2 kHz or 0.133~km~s$^{-1}$, is confirmed
by quantum chemical calculations which give a separation of 39.8 kHz and line strength
ratio 3:1 when H and $^{13}$C nuclear spin-rotation and spin-spin coupling between both nuclei are taken into account. We improve the spectroscopic constants of H$^{13}$CO$^{+}$ and determine the hitherto uncertain frequencies of its low-$J$ spectrum
to better precision by analysing the dark cloud L\,1512.
Attention is drawn to potentially high optical depths (3 to 5 in L\,1512) in quiescent
clouds, and examples are given for the need to consider the (1~--~0) line's doublet
nature when comparing to other molecular species, redirecting or
reversing conclusions arrived at previously by single-component interpretations.\\
We further confirm the hyperfine splitting in the (1~--~0) rotational
transition of $^{13}$CO that had already been theoretically predicted, and measured in the laboratory, to be of width about 46 kHz or, again,
0.13~km~s$^{-1}$. By applying hyperfine analysis to the extensive data set of the first IRAM key-project we show that $^{13}$CO optical depths can as for H$^{13}$CO$^+$ be estimated
in narrow linewidth regions
without recourse to other transitions nor to assumptions on beam filling factors,
and linewidth and velocity determinations can be improved.
Thus, for the core of L\,1512 we find an inverse proportionality between linewidth and
column density, resp. linewidth and square root of optical depth, and a systematic inside-out increase of excitation temperature and of
the $^{13}$CO:C$^{18}$O abundance ratio. Overall
motion toward the innermost region is suggested.
}

\titlerunning{
Hyperfine Structure in H$^{13}$CO$^{+}$ and $^{13}$CO}
\authorrunning{Schmid-Burgk et al.}

\maketitle

%________________________________________________________________

\section{Introduction}

It has long been recognised that cold interstellar clouds are excellent
laboratories for determining basic physical quantities of molecular structure. These clouds
also provide the only access to molecular species that are too unstable to
permit sufficient terrestrial production for in-depth investigation.
A good case in point here is the X-ogen of the early 1970s, alias oxomethylium or the formyl cation, HCO$^{+}$
(e.g.~Snyder et al. 1976) which has since gained prime importance for
studying the interactions between interstellar gas and magnetic fields.

In some of these clouds, microwave emission profiles are so narrow that line frequencies
can be determined to between seven and eight significant digits, relative to
other lines, corresponding to an accuracy of some ten meters per second in
velocity, or Mach 0.02 at typical temperatures.
A recent example is the re-analysis of the N$_2$H$^+$(1~--~0) transitions (Caselli et al. 1995) which led to a frequency determination 
with precision 7 kHz, difficult to attain with laboratory spectroscopy.
It thus becomes possible
to study minute motions in such objects that may be on the brink of
collapsing toward star formation.

With this aim we have observed the dark cloud Lynds 1512 in
numerous molecular transitions.
This object is a small accumulation of opaque blobs on the POSS plates,
situated near the edge of the Taurus-Auriga-Perseus complex. Ever since Myers \&
Benson (1983) discovered ammonia radiation from one of these blobs, this ``core'' has been known as
a source of particularly narrow emission lines and has consequently attracted
much attention, among many others that of Fiebig (1990) who from NH$_3$ observations derived a kinetic temperature of 9.7~K at the center,
Muders (1995) who drew attention to the cloud's well-structured velocity field, and more recently from
the extensive IRAM Key Project on pre-star-forming regions (Falgarone et al. 1998).

Somewhat to the East of the L\,1512 ammonia peak there exists a region particularly well-suited
for high-precision line studies because there, several tracers of high gas density like CCS,
HC$_{3}$N and C$_4$H show very narrow emission profiles whose widths indicate the kinetic temperatures of
the gas to be not more than 9 to 10~K.
There is a certain velocity gradient all over L\,1512 which
must be subtracted off when determining intrinsic linewidths,
but many coexisting species are subject to this same gradient and can therefore
be compared to each other very well
over the whole region.
Their intrinsic profiles then turn out to be basically identical.

There is one high-density tracer however which does not at all fit this general picture of uniformity:
H$^{13}$CO$^{+}$. We found its linewidth to be considerably broader than
that of the other species, and its profiles to be non-Gaussian all over this region.
Since HCO$^{+}$, being tied by its electric charge to collapse-retarding magnetic
field lines, could be the prime informer on the interaction between neutral and
charged molecules in dynamic clouds, it
seemed important to clarify the reason for this difference. We thus realised that the
magnetic moment of the $^{13}$C nucleus -- in conjunction with the much
less efficient moment of the single proton -- might cause an observable hyperfine splitting in the H$^{13}$CO$^{+}$(1~--~0)
line, of sufficient separation to be developed into a useful tool for spectral
analysis. Subsequently we searched for a similar splitting in
the (1~--~0) line of
the low-density tracer molecule $^{13}$CO as well. Below
we will report on some possibilities for line interpretation that emerge from the fact
that again the splitting is wide enough to be observed in quiescent cold clouds.

This paper is organised as follows. In section 2 we summarise the observations performed at
various telescopes, then present some general results in section 3.
Section 4 discusses the observed H$^{13}$CO$^{+}$(1~--~0) line shapes,
the determination of the width of the hyperfine splitting,
and the ensuing high optical depths of this isotopomer. In section 5 our quantum chemical
calculations concerning the hyperfine structure of this molecule are reported and improved spectroscopic constants derived.
Line frequencies for the low-$J$ rotational spectra of H$^{13}$CO$^{+}$ (and
HC$^{18}$O$^{+}$) can then be determined with high precision in section 6. Some examples of the analytical
possibilities provided by the seeming complication due to hyperfine structure
are given in section 7. Section 8 investigates the
corresponding splitting for the $^{13}$CO molecule, which allows several physical
trends to be established for the core of L\,1512. Finally, section 9 presents our conclusions.

\section{Observations}

\subsection{IRAM 30m Observations}

Our H$^{13}$CO$^+$(1~--~0) and HC$_3$N(10~--~9) data were taken simultaneously with the IRAM 30-meter telescope
on Pico Veleta during several sessions
between November 2000 and the summer of 2001. We used one of the 3mm SIS facility receivers. At the
line frequency of 86.754\footnote{Accurate frequencies for this and other lines are presented in Table 3}~GHz the
single sideband receiver noise temperature was $\sim$60~K and
the FWHM beam size 29$''$.
The spectra were analysed with the autocorrelator at 10~kHz (i.e. 0.034~km~s$^{-1}$) resolution.
We checked the pointing accuracy regularly on nearby continuum sources.
It was better than 2$''$.

The data were observed using the frequency switching technique with a shift
of $\pm$1~MHz, which is large enough to separate the very narrow lines
and small enough to get a very good baseline. The spectra were scaled to
the main beam brightness temperature scale using the efficiencies supplied
by the observatory ($\eta_{\rm mb}=0.78, \eta_{\rm ff}=0.92$).
Throughout the sessions the average system
noise temperatures including the atmosphere were $\sim$170~K (main beam scale)
which corresponds to an atmospheric opacity of about 0.1.

\subsection{HHT Observations}

The HCO$^+$(3~--~2) data were measured with the 10-meter
Heinrich Hertz Telescope (HHT) on Mt.\ Graham, AZ {\footnote {The 
HHT is operated by the Submillimeter Telescope Observatory on behalf of 
the Max-Planck-Institut f\"{u}r Radioastronomie and Steward Observatory of 
the University of Arizona.}} in February 1999 and March 2000. The facility
1mm SIS receiver operated in double sideband mode and had a receiver
noise temperature of $\sim$170~K. We used the 62.5~kHz resolution
filterbank to analyse the spectra.

At the line frequency of 267.558~GHz, the FWHM
beam size was 32$''$. The main beam efficiency was 0.78 and the forward
efficiency 0.95.
The pointing accuracy, from measurements of planets and other strong
continuum sources, was better than 3$''$.

In February 1999
the observing conditions were good and stable throughout the
entire observing period with average DSB system temperatures of
$\sim 500$~K (main beam scale),
corresponding to atmospheric opacities of $\sim 0.2$.
The conditions in March 2000 were worse, with average system temperatures
of $\sim 900$~K and opacities of $\sim 0.4$.

We used the
On-The-Fly mapping method to obtain fully sampled maps of the HCO$^+$(3~--~2)
emission toward L\,1512.

\subsection{Effelsberg 100m Observations}

The 30 GHz cooled HEMT receiver on the MPIfR 100-m telescope was used between January 1997
and March 1999 for mapping the HC$_3$N(3~--~2) triplet as well as transitions of some other species (see Fig. 2) in L\,1512. At the line frequency
of 27.294 GHz the FWHM beam size was 33$''$ and the receiver noise temperature $\sim 65$~K.
We employed the facility autocorrelators at a resolution of 1.5 and 2.5 kHz (0.015 and 0.025~km~s$^{-1}$), taking
special care to keep the velocity calibration always consistent.
On-The-Fly mapping was done in frequency-switching mode at a throw of $\pm$0.4 MHz.

The NH$_3$(1,1) emission
at 23.694~GHz
was measured repeatedly, using the old Effelsberg 1.3 cm K-band
as well as the new 1.3 cm HEMT receiver. The FWHM beam size at that
frequency is 39.6$''$. The receiver noise temperatures were typically of
order 50 K; main beam efficiency here is 58 percent. We mostly used
spectral resolutions of 1.25 or 2.5 kHz which correspond to 0.016 resp. 0.032 km~s$^{-1}$,
and observed in frequency-switching mode of throw $\pm$0.14 MHz.

\subsection{The $^{13}$CO(1~--~0) and C$^{18}$O(1~-~0) Spectra}

Apart from a few C$^{18}$O(1~--~0) spectra that were taken simultaneously with our
H$^{13}$CO$^+$(1~--~0) measurements mentioned above, all CO data used in this paper
come from the IRAM Key Project (IKP) on small-scale structure of pre-star-forming
regions whose observational conditions have been fully reported by Falgarone
et al. (1998) and will not be repeated here.

\section{
Results}

\begin{figure}[t]
\centering
\resizebox{10.5cm}{!}{\rotatebox[origin=br]{-90}{\includegraphics{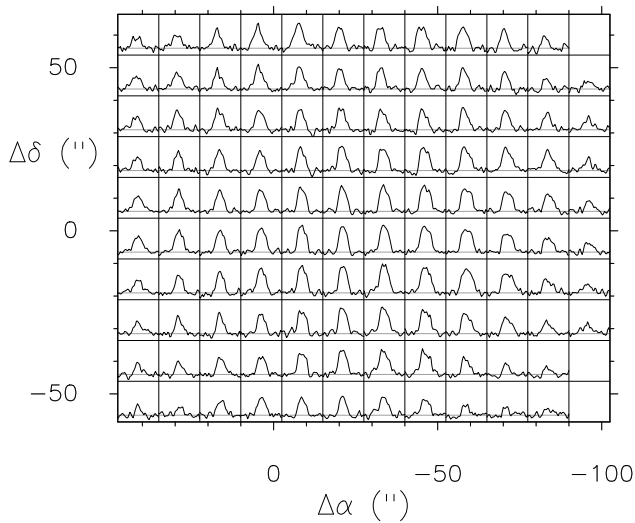}}}
\caption{
Grid of H$^{13}$CO$^+$(1~--~0) emission profiles from L\,1512.
T$_{mb}$ (from -0.3 to 1.5~K) is shown as function of v$_{LSR}$ (from 6.5 to 7.8~km~s$^{-1}$).
Offsets are relative to RA(1950) 05 00 54.4, Dec (1950) 32 39~00.
\label{fig1}}
\end{figure}

Our map of H$^{13}$CO$^+$(1~--~0) profiles is displayed in Fig.~1 where coordinate offsets are counted from
the position
RA(1950) 05 00 54.4, Dec (1950) 32 39 00.
Note that the true NH$_3$ and N$_2$H$^+$ peaks (i.e. the cloud's core center) are at
the approximate offset (-25$''$,10$''$).

Sizeable radiation extends somewhat further out than that of the other carbon-containing
high density tracers mentioned in the introduction. In contrast to these tracers H$^{13}$CO$^+$(1~--~0) does not show 
any brightness decline around the true ammonia peak that would signal molecular depletion
in the innermost region due to freeze-out onto grains. 
%Since the pregenitor of HCO$^+$,
%carbon monoxide, does seem to be centrally depleted,
This poses a problem. Absence of a
noticeable depletion dip is verified by the recently published H$^{13}$CO$^+$(1~--~0) map of Hirota et al.
(2003) which at a velocity resolution of 0.13~km~s$^{-1}$ covers the same area as ours plus an extension to the North where emission
{\it is} seen far beyond that from, e.g., HC$_3$N.

\begin{figure}[t]
\centering
\resizebox{9.5cm}{!}{\rotatebox[origin=br]{-90}{\includegraphics{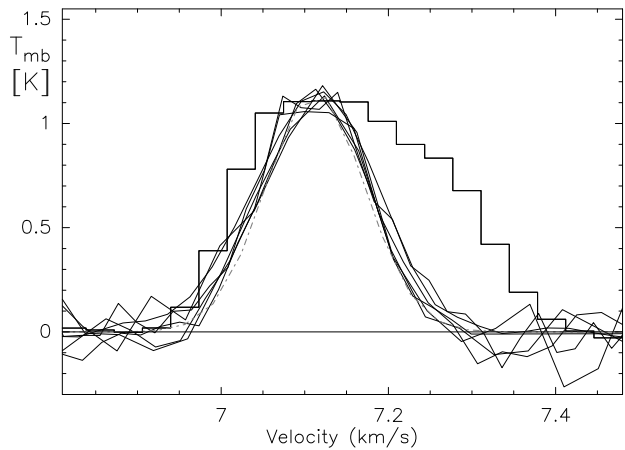}}}
\caption{
Line profiles  of H$^{13}$CO$^+$(1~--~0) (histogram),
NH$_{3}$(1,1:L1), C$^{18}$O(1~--~0), C$_{4}$H(3~--~2:J--7/2--5/2 F=4--3 and 5/2--3/2 F=2~--~1), HC$_{3}$N(3~--~2, F=2~--~1) and CCS(3~--~2),
all taken at offset (0$''$, -9$''$) and frequency-shifted for optimal overlap. The dash-dotted curve is the {\it reduced}
H$^{13}$CO$^+$(1~--~0) profile.
Scales are for H$^{13}$CO$^+$(1~--~0); rescaling for the other species as described in the text.
\label{fig2}}
\end{figure}

Both widths and shapes of our H$^{13}$CO$^+$(1~--~0) profiles differ noticeably from those of other
coexisting trace species. Typical FWHM widths in Fig.~1 are around 0.28~km~s$^{-1}$, with
a slight increase near offset (-60$''$, -10$''$) which is, however, less
marked than for numerous other molecules.
A sample spectrum, taken in long integration at offset (0$''$, -9$''$), i.e. about
halfway between the center and the edge of the high-density core of L\,1512,
is shown in Fig.~2 together with low-transition profiles of other coexisting species, namely
NH$_{3}$, C$^{18}$O, C$_{4}$H, HC$_{3}$N and CCS,
all from the same position.
The frequencies of these species here have been shifted and their amplitudes normalised for
better comparison; furthermore, each species' velocity scale 
has been expanded or compressed with the square root of the mass ratio between the molecule in question and H$^{13}$CO$^{+}$.
In this way it becomes evident that, at this position, their true linewidths go with (molecular mass)$^{-1/2}$; if all these species had atomic weight
30 amu, each line would be between 0.15 and 0.16~km~s$^{-1}$ wide. This value would have to be reduced for
several reasons like nonzero optical depth (typically a few to 15 percent reduction), velocity
gradient in the beam (a few percent for the velocity field of L\,1512)
plus likely also one along the line of sight, binning and folding of frequency-switched lines (another 5 percent), in order to obtain the intrinsic width. The purely thermal value for 10~K would be 0.126 km~s$^{-1}$.
Hence the gas is very cold and quiescent even though its position is not at
the core's center where the lowest temperatures are believed to exist.
There is no indication of macroscopic motion in any of the line shapes with the possible exception of
that of H$^{13}$CO$^+$.

The asymmetry of the stepped profile of Fig.~2 is not seen in any other comparable molecule
in this source.
It is not confined to a limited location.
At the majority of positions in Fig.~1 the lines have a steeper slope on the left than on
the right, never the other way round; the higher-velocity side often appears somewhat ``bumpy''.
This singular feature will be discussed next.

\section{
Interpretation of H$^{13}$CO$^{+}$(1~--~0) line shapes in L\,1512}

Line asymmetries can be caused by several mechanisms, like a sufficiently large velocity
gradient within the cloud, superposition of several emitters along the line of sight, or
(self)absorption by cooler gas on the near side of the source.
For the core of L\,1512 several other species suggest that in low-lying transitions neither of these
mechanisms is generally at work.
Since this core is known to exhibit very strong self-absorption in the HCO$^{+}$(3~--~2)
transition, however (cf Fig.~4b), it would not be surprising to find the third mechanism relevant in
H$^{13}$CO$^{+}$(1~--~0).
Unfortunately, the frequency of H$^{13}$CO$^{+}$(1~--~0) was not known a priori with high enough precision 
to let comparison with other lines decide the question of foreground absorption.
The absorption trough in the
main isotopomer's (3~--~2) transition clearly shifts across the line with position on the sky while the H$^{13}$CO$^{+}$ bump is static: any hypothetical
foreground absorbers of H$^{13}$CO$^{+}$(1~--~0) should thus be fixed locally to the emitters in velocity space, a somewhat special situation.
For these reasons we were led to suspect a hyperfine splitting (hfs) as the cause of the relatively large linewidths and asymmetry.

\subsection{Observation of the Hyperfine Splitting}

\begin{figure}[t]
\centering
\resizebox{9.44cm}{!}{\rotatebox[origin=br]{-90}{\includegraphics{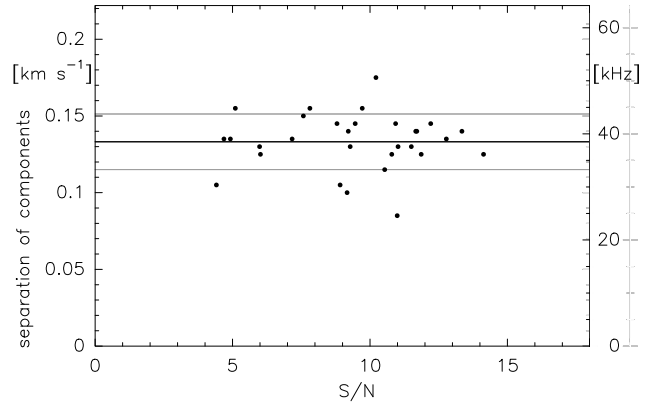}}}
\caption{
Best-fit separations of the primary and secondary components of the H$^{13}$CO$^{+}$(1~--~0) hyperfine doublet
for all independent measurements, as a function of the measurement's signal-to-noise ratio. S is mean
intensity over the line, N the noise at spectral resolution 0.0337~km~s$^{-1}$.
Horizontal lines give the mean value and rms error.
\label{fig3}}
\end{figure} 

\begin{figure}[t]
\centering
\resizebox{16.6cm}{!}{\rotatebox[origin=br]{-90}{\includegraphics{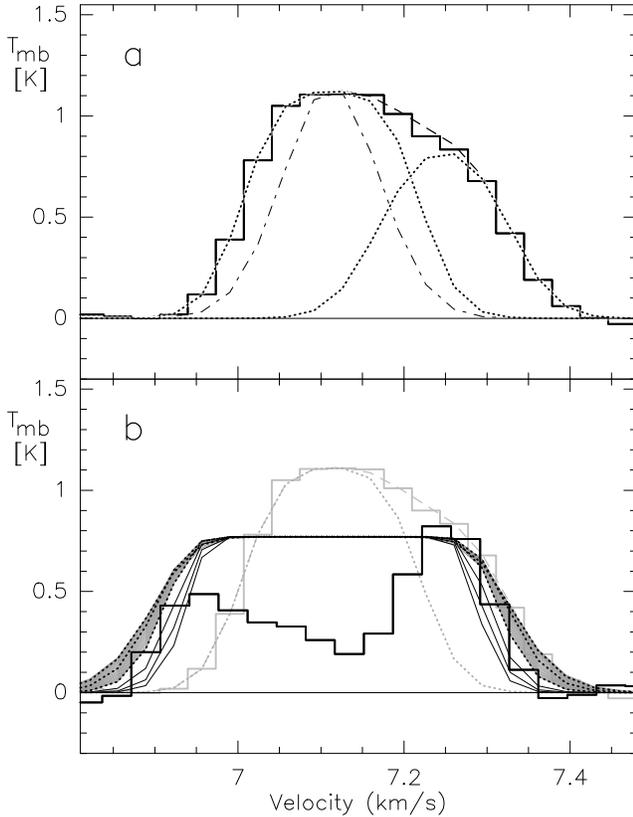}}}
\caption{
{\bf a)} Deep integration H$^{13}$CO$^{+}$(1~--~0) spectrum (histogram) from offset (0$''$, -9$''$), and its
resolution into primary and secondary hyperfine components (dots) which combine into the dashed profile. The dash-dot curve is the {\it intrinsic}
primary profile (i.e. before widening by optical depth).
{\bf b)} H$^{\bf 12}$CO$^{+}$(3~--~2) spectrum (dark histogram) from the same position, compared to
what would be expected when extrapolating, under simplifying assumptions, from the H$^{13}$CO$^{+}$(1~--~0) observation. Depending on the
latter's assumed total peak optical depth $\tau_{13}$, here 0.1, 1, 2, 3, 4, 5, the smooth flat-topped profiles would be expected for H$^{12}$CO$^{+}$(3~--~2).
Grey area: $\tau_{13}$ below 2; inner three full lines: $\tau_{13}$ = 3, 4, 5.
\label{fig4}}
\end{figure}

The magnetic moments of proton and $^{13}$C nucleus must in principle lead to such a
splitting, of unknown width. Therefore we reduced each profile, in the
$60''\times 110''$ field centered on $(0'',~0'')$
which contains the sources of narrowest line emission in the whole cloud,
by a set of assumed values
of the splitting width, keeping the line strength ratio fixed at 3:1 as prescribed by basic quantum theory (see section 5), and chose that value which optimally fits the measurement to be the 
profile's most likely value. All of these most likely values together are shown in Fig.~3; their consistency
leaves little doubt that the width of the splitting can well be determined in this region of
exceptional immobility and low temperature,
and rules out self-absorption as an essential feature.

The mean value of the hyperfine splitting so determined emerges as 0.133~km~s$^{-1}$ or
38.5 kHz, with a rms deviation of 0.018~km~s$^{-1}$. Since there are some 20 independent measurements
contained in the data of Fig.~3, the error might be considerably smaller still but, at
these levels, systematic errors may come into play.

With this result, and a line strength ratio of 3:1 (high:low frequency) as given by
theory (see below), we can resolve the H$^{13}$CO$^{+}$(1~--~0) profile of Fig.~2 into the two components
depicted in Fig.~4a by the dotted lines. Total optical depth, $\tau_{13}$, at this particular position is determined by comparing these components: it is
between around 3 and 5. Hence the stronger component must be considerably wider than the profiles of
the other molecular species shown in Fig.~2 whose $\tau$ is typically at or below unity. The {\it reduced} width (i.e. after correction for optical depth) of this H$^{13}$CO$^{+}$(1~--~0) profile turns out to be 0.138 $\pm$ 0.006~km~s$^{-1}$, however,
in agreement with the others' values mentioned above.
From $\tau_{13}$ and T$_{mb}$ the excitation temperature of the line follows as between 3.9 and 4.6~K, uncertain mainly because of uncertain calibration.

\subsection{Large optical depths of H$^{13}$CO$^{+}$ ?}

Before turning to theoretical calculations of the multiplet structure we make some
comments on this seemingly very high value of optical depth.
If confirmed, this could explain the seeming absence of central depletion that was mentioned above.
Virtually most published observations of H$^{13}$CO$^{+}$(1~--~0) to date have been
interpreted under the assumption of the line being optically thin,
with typical $\tau$ values around 0.1 to 0.4, although observed radiation temperatures
would then often require a questionably high excitation of the molecule. As will be
discussed below, the C$^{18}$O(1~--~0) emission from the position of Fig.~2 has an optical depth 
at line center of 0.72 and CO excitation temperature 9~K. We can then
ask what the corresponding optical depth of our H$^{13}$CO$^{+}$ line should be in case the
standard abundance ratios for $^{12}$CO:H$^{12}$CO$^{+}$ of $1.0\times 10^{4}$
for dark clouds (Irvine et al. 1987) and $\approx 7$ for $^{13}$CO:C$^{18}$O were applicable in the present context. With the two excitation
temperatures given above, we calculate 44 percent of all C$^{18}$O
molecules (in LTE), and about the same percentage of all H$^{13}$CO$^{+}$ (not in LTE) to be
in state $J$=1. The $\tau({\rm H^{13}CO^{+}}):\tau({\rm C^{18}O})$ ratio then is
2.24$\times7\times10^{-4}\times
(\mu({\rm H^{13}CO^{+}}):\mu({\rm C^{18}O}))^{2}$,
where the factor 2.24 contains all the statistics (and whose narrow error margin of
10 to 15 percent is hardly negotiable);
$\mu$ is the electric
dipole moment of the species in question. The $\mu$ quotient is $\approx$ 35 if $\mu({\rm H^{13}CO^{+}})$ = 3.93 D
(Botschwina et al. 1993) is assumed; it could be somewhat higher.

In this way we deduce the H$^{13}$CO$^{+}$ line's optical depth to be about twice that of C$^{18}$O,
or 1.5 at the present position, a value by no means negligible and quite compatible
with the 3 to 5 derived above considering the several uncertainties involved.
The most insecure factor would seem to be the CO:HCO$^{+}$ ratio, a reduction of which
by fifty percent
could bring our estimates nicely in line.

Further support for high H$^{13}$CO$^{+}$ optical depths comes from a somewhat simplistic comparison
to the main isotopomer's (3~--~2) emission from the same position on the sky (the dark histogram in Fig.~4b). The lower $\tau_{13}$, the higher the intrinsic linewidth must be to 
fit the H$^{13}$CO$^{+}$ profile. This means higher H$^{12}$CO$^{+}$ linewidth as well, so that a certain $\tau_{13}$
limit is given by the measured H$^{12}$CO$^{+}$ width.
It is assumed here that both profiles originate from physically similar gas.
If both isotopomers
had T$_{ex}$ values around 9~K, $\tau_{13}$ would have to be at least 0.17
to produce the amplitude of the spectrum of Fig.~2.
Hence the optical depth of H$^{12}$CO$^{+}$(3~--~2) would be some 1.24$\times$abundance ratio, or $\approx$ 50 to 90 times as high,
i.e. around 10 to 15. Combined with the required intrinsic linewidth this results in a
minimum width of the H$^{12}$CO$^{+}$ profile that is considerably wider than what is observed.
How would this change in the case of non-LTE?
Under the two natural assumptions that excitation
temperatures T$_{ex}$ for the rarer isotopomer are not higher than for the dominant one
(because radiative excitation will push the optically deeper species closer to LTE),
and that T$_{ex}$ does not increase with increasing rotational transition,
it is straightforward to determine {\it lower} limits on the $\tau_{12}:\tau_{13}$ ratio {\it independent} of the
degree of deviation from LTE. For each assumed value of $\tau_{13}$ and its
corresponding intrinsic linewidth, a minimum width of the H$^{12}$CO$^{+}$ profile can thus be
computed. Examples for various $\tau_{13}$ are given in Fig.~4b as the smooth and dotted dark curves, using a $^{12}$CO:$^{13}$CO abundance ratio of 70 for demonstration.
As $\tau_{13}$
is increased, the computed H$^{12}$CO$^{+}$ profiles contract. The grey-shaded region covers $\tau_{13}$
from 0.1 to 1 to 2 (dots); the full lines correspond to $\tau_{13}$ = 3, 4, and 5 (the innermost one).

It seems that $\tau_{13}$ values below about 2 would require too wide an H$^{12}$CO$^{+}$ line. Due to the
strong selfabsorption of that line the argument depends somewhat on its true
(unabsorbed) value of T$_{mb}$. From comparison with the other positions in the map (where
selfabsorption shifts widely across the line), T$_{mb}$ values below 0.75~K are
improbable. 0.75~K means T$_{ex}$(H$^{12}$CO$^{+}$(3~--~2)) equal to 4.60~K.
Note that nearly all across the map H$^{12}$CO$^{+}$ widths are too narrow to allow $\tau_{13}$ values much below about 2.

Since the width of a double-structure profile may considerably exceed the
intrinsic linewidth, the usual formulae for column density have to be
reconsidered. If the molecules have a Gaussian distribution in velocity
space, with the full width at half maximum equal to $\Delta v$, the general expression for
the column density of the ground state, $N_{0}$, is,
independent of hyperfine multiplicity and for arbitrary optical depths,
$$N_{0} = c_0 \cdot (\nu_{10}/c)^{3} \cdot A_{10}^{-1}\cdot f_{bf}\cdot(1-e^{-T_{0}/T_{ex}})^{-1} \cdot\tau_{13}\cdot\Delta v$$\\
where $c_0 = \frac{4}{3} (\pi^{3}/{\rm ln}(2))^{1/2}$,
$\tau_{13}$ the {\it total} peak optical depth, i.e. the sum of the two components,
$A_{10}$ the Einstein coefficient for the 1~--~0 transition, $\nu_{10}$ the line frequency,
T$_{0} = h\nu_{10}/k$ = 4.16~K, T$_{ex}$ the excitation temperature between states 1 and 0, and $f_{bf}$ the beam filling factor.
$\Delta v$ is the {\it intrinsic} linewidth of either component, before broadening by
optical depth, and {\it not} the observed profile width.
$A_{10}$ is $3.9\times10^{-5}\cdot (\mu/3.93\;D)^{2}\;{\rm s}^{-1}$, hence 
$$N_{0} = 5.50\times10^{11}\cdot\tau_{13}\cdot\Delta v \cdot (1-e^{-T_{0}/T_{ex}})^{-1}  {\rm cm}^{-2}$$\\
if $\mu$ = 3.93 D and $f_{bf}$ = 1 is assumed and $\Delta v$ is measured in~km~s$^{-1}$.

To translate this ground-state into total column density, the
partition function is required. In the present case, where T$_{ex}$ is around
4 to 4.6~K, this can be obtained quite reliably since levels above $J$=1 are
hardly occupied, contributing altogether less than 20 percent to the partition function.
That function is then 2.5 to within less than 15 percent uncertainty, so that the total
H$^{13}$CO$^{+}$ column density at the position of Fig.~2 becomes $1.2\times10^{12}$ cm$^{-2}$ for the
T$_{ex}$, $\tau_{13}$, and $\Delta v$ values derived above.

\section{
Hyperfine structure in H$^{13}$CO$^+$ and related species}

The presence of one or more nuclei with spins of $I \geq 1$ causes
a rotational transition to appear as a group of several components due
to interactions between the electric quadrupole moments of these
nuclei and the electric field gradients present around them.
The splitting between the lines is frequently resolved in laboratory
spectroscopy, and to some extent even in astronomical observations, in
particular for transitions with low values of $J$.

It is less well known that for molecules with one or more nuclei having
$I \geq 1/2$ there exists another hyperfine effect: the coupling of the nuclear magnetic
spin to rotation. While in the case of nuclei with $I \geq 1$ this
effect results only in a shift of the quadrupole components, it causes
each rotational level with $J > 0$ to split into two for nuclei with
$I = 1/2$. In the laboratory, the nuclear spin-rotation splitting due to a nucleus with
$I = 1/2$ is generally not resolvable by Doppler-limited rotational
spectroscopy. Sub-Doppler methods, such as Lamb-dip spectroscopy or
microwave Fourier transform spectroscopy, may overcome this obstacle.
%% two exemplary references may be included

The nuclear spin-rotation coupling constants, $C$, describing this
effect are proportional to the rotational constants,
to the magnetic moments and to the inverse of the spin-multiplicity
$2I + 1$ of the nuclei. In addition, the presence of low-lying
electronic states causes the magnitude of the $C$s to increase
in a complex way.

The nuclear spin-rotation coupling constants for H$^{13}$CO$^+$ have
not yet been determined experimentally. Therefore, quantum chemical
calculations were performed at the ETH Z\"urich in order to
evaluate the nuclear spin-rotation coupling constants of several
isotopomers of HCO$^+$. In addition, similar calculations were
carried out for isotopomers of the isoelectronic molecule HCN and for the related
CO molecule, for both of which experimental results are available, in order to
check the reliability of the quantum chemical calculations.

The {\it ab initio} calculations were performed at the experimental
$r_{e}$ molecular geometries of
HCO$^+$ (Bogey et al. 1981), HCN (Maki et al. 2000), and CO (Coxon \& Hajigeorgiou 1992)
using the \textsc{Dalton} program (Helgakar et al. 1997).
Restricted Active Space (RAS) wavefunctions and the
Multi-Configuration Self Consistent Field (MCSCF) method were used;
single and double excitations from the fully active (RAS2) space to
the limited entry (RAS3) space were considered.  The orbital spaces
used were chosen based on a consideration of the MP2 Natural Orbital
occupation numbers (Jensen et al. 1988a, 1988b).

Several different calculations were done in order to test the
influence of basis set and orbital space size on the calculated
results.  It was found that the results were well converged with
respect to the different basis sets used (cc-pVXZ (X=D,T,Q),
aug-cc-pVXZ (X=D,T,Q), cc-pCVDZ, and aug-cc-pCVXZ (X=D,T)), and that
expansion of the secondary space beyond a certain limit resulted in a
larger calculation but did not yield a more accurate result.
Although the various orbital spaces tested were sometimes slightly
different for each of the molecular species studied, several of them were
common to all three.  The final calculations were done using a fairly
large orbital space common to all three molecules.  The results are given
in Table 1; these were obtained from calculations done using the aug-cc-pCVTZ
basis set and 2 inactive, 5 fully active, and 18 limited-entry
orbitals.

\begin{table*}
\caption{Nuclear spin-rotation constants (kHz)$^{a}$ for HCO$^+$
and related molecules, calculated with the aug-cc-pCVTZ basis set,
in comparison to experimental results (in {\it italics}). \label{fittare}}
\begin{center}
\begin{tabular}{|l r@{}l  r@{}l  r@{}l  r@{}l|}
\hline
&\multicolumn{2}{c}{$C$(H and D)}&\multicolumn{2}{c}{$C$($^{13}$C)}&\multicolumn{2}{c}{$C$($^{17}$O)}&\multicolumn{2}{c|}{$C$($^{14}$N)}\\
\hline
 & & & & & & & & \\
H$^{13}$CO$^+$ &  --5.55 &/ ---             & 19.4 &/ {\it 18.9\,(39)}$^b$            &    --- &/ ---              & --- &/ ---            \\
HCO$^+$        &  --5.70 &/ ---             &  --- &/ ---             &    --- &/ ---              & --- &/ ---            \\
HC$^{18}$O$^+$ &  --5.45 &/ ---             &  --- &/ ---             &    --- &/ ---              & --- &/ ---            \\
HC$^{17}$O$^+$ &  --5.57 &/ ---             &  --- &/ ---             & --21.6 &/ {\it --19.5\,(16)}$^c$ & --- &/ ---            \\
HCN            &  --4.80 &/ {--\it 4.35\,(5)}$^d$ &  --- &/ ---             &    --- &/ ---              & 9.9 &/ {\it 10.13\,(2)}$^d$ \\
D$^{13}$CN     &  --0.59 &/ {\it --0.6\,(3)}$^e$  & 14.3 &/ {\it 15.0\,(10)}$^e$  &    --- &/ ---              & --- &/ ---            \\
$^{13}$CO      &     --- &/ ---             & 32.4 &/ {\it 32.63\,(10)}$^f$ &    --- &/ ---              & --- &/ ---            \\
C$^{17}$O      &     --- &/ ---             &  --- &/ ---             & --30.7 &/ {\it --31.6\,(13)}$^g$ & --- &/ ---            \\
 & & & & & & & & \\
\hline
\end{tabular}
\end{center}

\fussy $^{a}$ Numbers in parentheses are one standard deviation in units of the least significant \nolinebreak digit

$^b$ this work
\hspace{1.5cm} $^d$ Ebenstein \& M\"{u}nter (1984)
\hspace{1cm} $^f$ Klapper et al. (2000)

$^c$ Dore et al. (2001)
\hspace{0.3cm} $^e$ Garvey \& De Lucia (1974)
\hspace{1.1cm} $^g$ Refit of Cazzoli et al. (2002b)

\end{table*}

The agreement between the calculated and experimental spin-rotation
constants of Table 1 is very good, their deviations frequently being within
the experimental uncertainties. Deviations outside the experimental
uncertainties may be caused by the neglect of vibrational effects
since the experimental data refer to the ground vibrational state
whereas the calculations refer to the equilibrium state.
In addition, limitations due to the basis sets and the method used may
lead to some deviation. But in most instances these deviations are
expected to be of the order a few percent.

In the case of HCO$^+$ and HC$^{18}$O$^+$, only the H nucleus has non-zero spin,
thus the angular momenta are coupled as

{\bf J} + {\bf I}$_H$ = {\bf F}.

Hence, the hyperfine splitting can be calculated in a straightforward manner.
In H$^{13}$CO$^+$, however, both the H and $^{13}$C nuclei have a
spin of 1/2. The spin-rotation constant of $^{13}$C is larger than that
of H, thus the angular momenta are coupled in the following way:

{\bf J} + {\bf I}$_C$ = {\bf F}$_C$;
{\bf F}$_C$ + {\bf I}$_H$ = {\bf F}.

In addition, the coupling between the H and $^{13}$C nuclei
have to be considered. For light nuclei such as H and $^{13}$C, the
indirect, electron coupled part is small for both the scalar and the
tensorial components. Therefore, these have been neglected in the present
calculation, so that the tensorial spin-spin coupling constant $S$ is
approximated by the direct part $d$, which was calculated from the
structure as $-25$ kHz. The resulting overall hyperfine shifts,
assignments and relative intensities of the individual components
of the 1 -- 0 transitions of H$^{13}$CO$^+$ and HC$^{18}$O$^+$
are given in Table 2.
\begin{table}

\caption{
Calculated hyperfine components of the
1~--~0 rotational transitions
of H$^{13}$CO$^+$ and HC$^{18}$O$^+$
using the nuclear spin-rotation constants of Table 1 and
the $^1$H-$^{13}$C spin-spin coupling constant
for H$^{13}$CO$^+$
(see text). \label{fittare}}
\begin{center}
\begin{tabular}{|c c r@{}l c|}
\hline
species        & assignment$^a$ & \multicolumn{2}{c}{shift (kHz)$^b$} & rel. intensity \\
\hline
   &   &   &   &      \\
H$^{13}$CO$^+$ & 0.5,1 -- 0.5,0 &   --29&.9          & 0.065 \\
               & 0.5,1 -- 0.5,1 &   --29&.9          & 0.185 \\
               & 1.5,2 -- 0.5,1 &      9&.4          & 0.417 \\
               & 1.5,1 -- 0.5,0 &     10&.4          & 0.185 \\
               & 1.5,1 -- 0.5,1 &     10&.4          & 0.065 \\
               & 0.5,0 -- 0.5,1 &     11&.2          & 0.083 \\
               &                &       &              &     \\
HC$^{18}$O$^+$ & 1.5 -- 0.5     &    --2&.7          & 0.667 \\
               & 0.5 -- 0.5     &      5&.4          & 0.333 \\

\hline
\end{tabular}
\end{center}

$^a$ $F_1',F' - F_1'',F''$

$^b$ With respect to hypothetical unsplit line frequency

\end{table}

At linewidths of more than about 2 kHz and substantially less than
40 kHz, the six hyperfine components of H$^{13}$CO$^+$(1~--~0) that have non-zero
intensity overlap partially,
in effect yielding {\it two} lines with the intensity ratio 1 : 3 and
separation of 39.8 kHz. Taking into account the neglect of vibrational
effects and the deficiencies in the calculational method and in the basis set, this
splitting is uncertain to at least 2 kHz, possibly to as much as 5 kHz.
Therefore, the agreement with the splitting of $38.5 \pm 5$ kHz that we derived above from
astronomical observations can be considered very good.

%% (Simulations ?? More discussion ?)

\section{Determination of precise line frequencies for H$^{13}$CO$^{+}$(1~--~0) and HC$^{18}$O$^{+}$(1~--~0)}

With this knowledge of hyperfine splitting we are now able
to considerably improve on H$^{13}$CO$^{+}$(1~--~0) frequency values published
previously. These are entered in Table 3 together with their
given errors; the latter amount to some 0.15~km~s$^{-1}$ in velocity uncertainty,
\begin{table*}
\caption{List of Frequencies. \label{fittare}}
\begin{center}

\begin{tabular}{|cc r@{}l p{4.3cm}|}
\hline
\multicolumn{1}{|c}{Molecule}
& \multicolumn{1}{c}{Transition}
& \multicolumn{2}{c}{Frequency (MHz)$^{a}$}
& \multicolumn{1}{l|}{Reference}
  \\ \hline
 & & & & \\
H$^{13}$CO$^{+}$ & 1--0$^{b}$ & {\sf 86754}&{\sf .2982(35)} & this work, ``primary'' \\ 
  & 1--0$^{b}$ & {\sf 86754}&{\sf .2589(69)} & this work, ``secondary'' \\
 & & & & \\
  &  {\sf mean} 1--0 & {\sf 86754}&{\sf .2884(46)} & this work \\
  & & {\it 86754}&{\it .329(39)} & Woods et al. (1981) \\
  &   & {\it 86754}&{\it .294(30)} & Gu\'{e}lin et al. (1982) \\
 & & & & \\
   & {\sf mean} 2--1 & {\sf 173506}&{\sf .697(10)}$^{c,d}$ & this work \\
   &      & {\it 173506}&{\it .782(80)} & Bogey et al. (1981) \\
 & & & & \\
  & {\sf mean} 3--2  & {\sf 260255}&{\sf .339(35)}$^{c,d}$ & this work \\
 & & {\it 260255}&{\it .478}$^{d}$ & Pickett et al. (1998) \\
  &   & {\it 260255}&{\it .339(35)} & Gregersen, Evans (2001) \\
 & & & & \\
  & {\sf mean} 4--3  & {\sf 346998}&{\sf .347(89)}$^{c,d}$ & this work \\
 & & {\it 346998}&{\it .540}$^{d}$ & Pickett et al. (1998) \\
 & & & & \\
HC$^{18}$O$^{+}$ & {\sf mean} 1--0$^{b}$ & {\sf 85162}&{\sf .2231(48)} & this work \\
                 &      & {\it 85162}&{\it .157(47)} & Woods et al. (1981) \\
 & & & & \\
\hline
 & & & & \\
$^{13}$CO & 1$_{1.5}$--0$_{0.5}$ & {\it 110201}&{\it .3703(2)}$^{d}$ & CDMS (2001) \\
$^{13}$CO & 1$_{1.5}$--0$_{0.5}$ & {\it 110201} & {\it .3697(10)} & Cazzoli et al. (2002a) \\
$^{13}$CO & 1$_{0.5}$--0$_{0.5}$ & {\it 110201}&{\it .3213(2)}$^{d}$ & CDMS (2001) \\
$^{13}$CO & 1$_{0.5}$--0$_{0.5}$ & {\it 110201}&{\it .3233(10)} & Cazzoli et al. (2002a) \\
 & & & \\
 &  {\it mean} 1--0 & {\it 110201}&{\it .3541(51)} & Winnewisser et al. (1985) \\
 & & & & \\
C$^{18}$O & 1--0 & {\it 109782}&{\it .1734(63)} & Winnewisser et al. (1985) \\
          &      & {\it 109782}&{\it .172(20)} & Klapper et al. (2001)\\
          &      & {\it 109782}&{\it .17569(40)} & Cazzoli et al. (2003)\\
 & & & & \\
HC$_{3}$N & 3$_{4}$--2$_{3}$ & {\it 27294}&{\it .3451(4)}$^{d}$ & CDMS (2001) \\
 & 3$_{3}$--2$_{2}$ & {\it 27294}&{\it .2927(3)}$^{d}$ & CDMS (2001) \\
 & 3$_{2}$--2$_{1}$ & {\it 27294}&{\it .0758(3)}$^{d}$ & CDMS (2001) \\
 & & & & \\
 &  {\it mean} 3--2 & {\it 27294}&{\it .289(10)} & Creswell et al. (1977) \\
 & & & & \\
 & 10$_{11}$--9$_{10}$ & {\it 90979}&{\it .0024(10)}$^{d}$ & CDMS (2001) \\
 & 10$_{10}$--9$_{9}$ & {\it 90978}&{\it .9948(10)}$^{d}$ & CDMS (2001) \\
 & 10$_{9}$--9$_{8}$ & {\it 90978}&{\it .9838(10)}$^{d}$ & CDMS (2001) \\
 & & & & \\
 &  {\it mean} 10--9 & {\it 90979}&{\it .023(20)} & Creswell et al. (1977) \\
 & & & & \\
\hline
\end{tabular}
\end{center}

\fussy $^{a}$ Numbers in parentheses are one standard deviation in units of the least significant \nolinebreak digit

$^{b}$ See Table 2 for detailed assignments % $^{p}$:``primary''; $^{s}$: ``secondary''

$^{c}$ These lines are multiplets of total spread (essentially) $\approx$20 kHz

$^{d}$ Frequencies {\it calculated} from the spectroscopic constants

\end{table*}
harmfully large for comparisons with other species
in cold clouds. Note that ignoring hyperfine structure must {\it necessarily} lead to an
intrinsic frequency uncertainty of 8 kHz due to
the shifting of the overall line center with optical depth.
Taking hyperfine structure into account, on the other hand, allows one to determine apparent
H$^{13}$CO$^{+}$(1~--~0) velocities in narrow-line regions such as the one
discussed here to much better precision than this.
If the gas velocities are known reliably from some other
measurements and are seen to apparently differ from H$^{13}$CO$^{+}$(1~--~0), the frequencies of the latter's doublet must be 
adjusted accordingly.

This can be done in our region in L\,1512 because this shows a remarkable similarity in the intensity,
velocity and linewidth distributions for a number of
molecular transitions such as the HC$_{3}$N (3~--~2) and (10~--~9) main
triplets, the C$_{4}$H(3~--~2) quadruplet, CCS $(3~-~2)$ and $(2~-~1)$ as well as
the H$^{13}$CO$^{+}$(1~--~0) doublet. Line widths are very close to thermal (at $\approx$ 9~K)
for all of these transitions, and $v_{LSR}$ maps are nearly
indistinguishable from species to species except for
constant $v_{LSR}$ offsets between them which must be the result of inaccuracies in their 
assumed rest frequencies.

Here we compare the
H$^{13}$CO$^{+}$(1~--~0) doublet transition to the two HC$_{3}$N triplets just mentioned
because for HC$_{3}$N, exceptionally precise rest frequencies have recently
been published by Thorwirth, M\"{u}ller and Winnewisser (2000), CDMS (2001); see Table 3. To a certain extent we can
verify their precision by comparing the HC$_{3}$N(3~--~2) velocities observed at the
100m telescope, over the same region as before plus an equally
large extension immediately to the West,
with 30m measurements of (10~-~9) over the same area.
Fig.~5 presents the results (filled squares) of this comparison, 
derived from some 20 truly independent (10~-~9) measurements that we obtained from the $\approx$
80 point maps by coadding four nearest neighbors.
The mean (10~-~9) to (3~-~2) velocity difference emerges as 0.0010 $\pm$ 0.0055~km~s$^{-1}$, or 0.3 $\pm$ 1.7 kHz, in
complete agreement with the CDMS error bars of 1.0 resp. 0.35 kHz.
Since we had verified the
frequency accuracy of the 100m system in front of the autocorrelator to be better than 1 Hz, by injecting a precisely
defined artificial signal,
we may therefore expect the 30m system to be of similar precision unless frequency prediction and
system inadequacy have conspired in a particularly misleading way.
\begin{figure}[t]
\centering
\resizebox{17.5cm}{!}{\rotatebox[origin=br]{-90}{\includegraphics{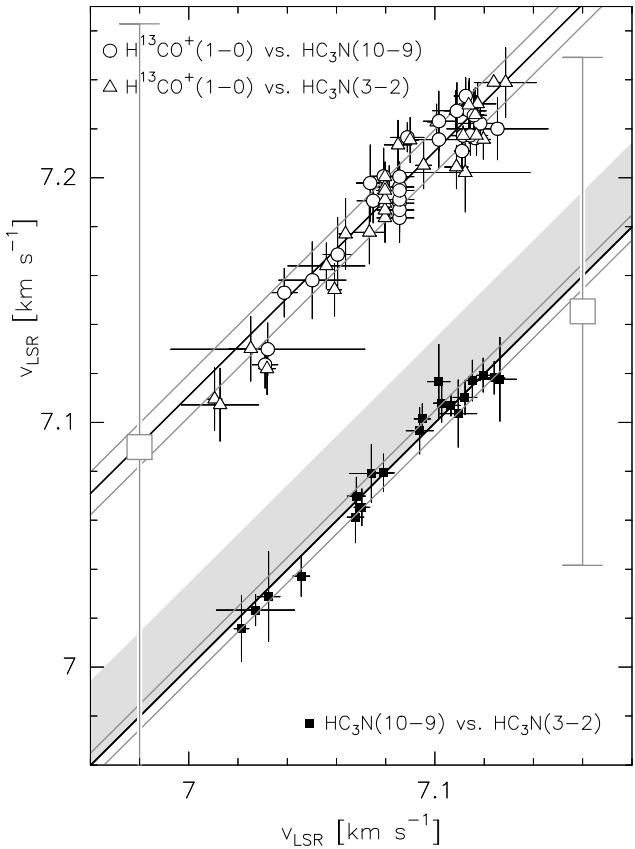}}}
\caption{
Correlations between the velocities of three different pairs of transitions, in the L\,1512 area described in the text. 
Filled squares: HC$_{3}$N(10~--~9) vs. HC$_{3}$N(3~--~2); open circles:
H$^{13}$CO$^{+}$(1~--~0) vs. HC$_{3}$N(10~--~9); open triangles: H$^{13}$CO$^{+}$(1~--~0) vs. HC$_{3}$N(3~--~2).
For demonstration we assume here
for the H$^{13}$CO$^{+}$(1~--~0) primary's frequency the Woods et al. value,
indicated on the left by the open square and its error bars, to be the correct one; the much more adequate
Gu\'{e}lin et al. frequency is correspondingly depicted by the open square on the right.
The dark diagonals represent the best linear fits of the correlations under the assumption
that velocity offsets between transitions are constant over the
whole field.
Fainter diagonals show the rms deviations from these best linear fits.
The shaded ribbon delineates the velocity uncertainty {\it necessarily} inherent in any
H$^{13}$CO$^{+}$(1~--~0) interpretation that employs the correct rest frequency but does not take hyperfine splitting into account; points of high optical depth accumulate near the ribbon's upper edge.
\label{fig5}}
\end{figure}

Our H$^{13}$CO$^{+}$(1~--~0) observations and their simultaneously measured HC$_{3}$N(10~--~9) counterparts
likewise show a clear correlation with each other, as demonstrated by the open circles of Fig.~5.
Here, we arbitrarily chose the Woods et al. (1981) frequency of 86754.329 MHz
for the primary component of H$^{13}$CO$^{+}$(1~--~0).
The corresponding H$^{13}$CO$^{+}$ correlation with HC$_3$N(3~--~2) (open triangles) is indistinguishably
similar to that with (10~--~9) as far as the mean velocity difference is concerned, 
so that we coadded the two correlations to arrive at a H$^{13}$CO$^{+}$(1~--~0) -- HC$_{3}$N
difference of the signal-to-noise-squared weighted mean velocities of
0.1066 $\pm$ 0.0120~km~s$^{-1}$ for 24 truly independent H$^{13}$CO$^{+}$(1~--~0) data points. This
corresponds to 30.8 $\pm$ 3.5 kHz by which the Woods et al. frequency must be lowered
in order to achieve velocity concordance between H$^{13}$CO$^{+}$ and HC$_{3}$N.
We thus arrive at a frequency of 86754.2982 $\pm$ .0035 MHz for the
primary line of the H$^{13}$CO$^{+}$(1~--~0) doublet, and of
86754.2589 $\pm$ .0069 MHz for the secondary. (The latter error is larger
due to the uncertainty in the precise magnitude of the hyperfine split).
The best value for {\it single-component} interpretations of the doublet would thus
be 86754.2884 MHz,
a mere 6 kHz or 0.02~km~s$^{-1}$ away from the result of
Gu\'{e}lin, Langer and Wilson(1982).
Note that all errors given here are (very conservatively) rms deviations from the mean;
for $\approx$20 independent data, the formal statistical error should be
considerably smaller still. But even these rms values improve on
published errors by factors of order 10, as shown by the large open squares in Fig.~5 which represent Woods et al.(1981) left, Gu\'{e}lin et al.(1982) right, both with
their error bars.

The apparent drop near the left end of the H$^{13}$CO$^{+}$(1~--~0) -- HC$_{3}$N correlation is not
indicative of a realistic degree of true uncertainties but has an astronomical
reason: these points all lie on the periphery of the cloud core where redshifted
emission from the core mixes with the blueshifted one from surrounding gas.
In these regions the ratio of blue to red contributions varies from species to species. For the mean
and rms values these points are, however, not of concern.

Based on our improved (1~--~0) frequencies,
on the (2~--~1) frequency from Bogey et al. (1981), and on the (3~--~2) frequency recently
proposed by Gregersen \& Evans (2001) we have redetermined the spectroscopic constants of H$^{13}$CO$^+$
which are presented in Table 4 together with their
previous values from Bogey et al. (1981). Our unsplit line frequencies
for the (2~--~1), (3~--~2) and (4~--~3) transitions, calculated with these new constants, are included in Table 3. Additional
predictions are available in the CDMS (M\"uller et al. 2001). Their hyperfine
pattern consists basically of two features separated by $\approx$ 20 kHz. With
increasing transition frequency such splitting becomes very hard to resolve because
of the increasing frequency width of the lines.

\begin{table}

\caption{Spectroscopic constants$^a$ (MHz) of H$^{13}$CO$^+$ as determined from the present
astronomical observations, compared with previous values$^b$. \label{fittare}}

\begin{center}
\begin{tabular}{|l r@{}l|}
\hline
parameter                       & \multicolumn{2}{c|}{value$^b$} \\
\hline 
 & & \\
$B$                             &   43377&.3011(27) \\
                                & {\it 43377}&{\it .320(40)}  \\
$D$ $\times$ 10$^{-3}$            &      78&.37(39)   \\
                                &   {\it 78}&{\it .3(75)}    \\
$C$($^{13}$C) $\times$ 10$^{-3}$ &      18&.9(39)      \\
$C$(H) $\times$ 10$^{-3}$        &     --5&.55$^c$     \\
$d$ $\times$ 10$^{-3}$           &    --25&.0$^c$      \\
 & & \\

\hline
\end{tabular}
\end{center}

$^a$ Numbers in parentheses are one standard deviation in units of the
least significant digit

$^b$ In {\it italics}; Bogey et al. (1981)

$^c$ Fixed to calculated value, see Table 1 and text

\end{table}

Both our (1~--~0) and (2~--~1) frequencies are at the lower end of the ranges given by Woods et al. (1981) resp. Bogey et al. (1981).
Since the (2~--~1) laboratory frequency contributes only little to our spectroscopic
constants (because of its large uncertainty), the very good agreement of our
(3~--~2) prediction with the astronomical value from Gregersen \& Evans (2001)
comes as no surprise.

Simultaneous measurements of HC$^{18}$O$^{+}$(1~--~0) and H$^{13}$CO$^{+}$(1~--~0) finally led to
a transition frequency of 85162.2231 $\pm$ 0.0048 MHz for the former, which itself
has hyperfine splitting on the order of 8 kHz with a line strength ratio of 1:2.
The {\it primary's} value thus is some 2.7 kHz lower; see Table 2.

\section{Can this hyperfine structure be of importance?}

One might think that a line splitting of not more than the thermal width of very
cold ($\approx$ 10~K) interstellar gas could hardly affect the
interpretation of measured line profiles. 
We now want to give some examples to the contrary.

\subsection{The case of L\,1544}

The first example for the potential usefulness of the doublet structure of H$^{13}$CO$^{+}$ concerns a recent
publication by Caselli et al. (2002).
These authors mapped a number of HCO$^{+}$ and other isotopomers in
the dense infalling core of L\,1544 and presented high resolution 
line profiles from its central dust emission peak.
In accordance with previous results (e.g.~Tafalla et al., 1998)
they found many optically thin lines to show double or highly
asymmetric profiles toward the dust peak. With self-absorption 
being unlikely for very thin lines, they argued for two separate velocity
components along the line of sight to the dust peak (here: ``blue'' and ``red'' for
short) which they discussed in the framework of the Ciolek \& Basu (2000) model of
an almost edge-on disk that is radially contracting; the central
region is thought to be depleted in CO-related species.
The authors noted the H$^{13}$CO$^{+}$(1~--~0) profile to be remarkably different from the
other five HCO$^{+}$ isotopomer transitions which they presented for
the dust peak. 
The former is much broader than the others, suggesting it to be
optically thick, and its blue and red velocity components are
separated by 0.36~km~s$^{-1}$, in contrast to the $\approx$ 0.24~km~s$^{-1}$ that hold
for {\it all} the others.

We display their (1~--~0) profiles of HC$^{18}$O$^{+}$ and H$^{13}$CO$^{+}$ in Figs. 6a and b (dark histograms).
\begin{figure}[t]
\centering
\resizebox{12.4cm}{!}{\rotatebox[origin=br]{-90}{\includegraphics{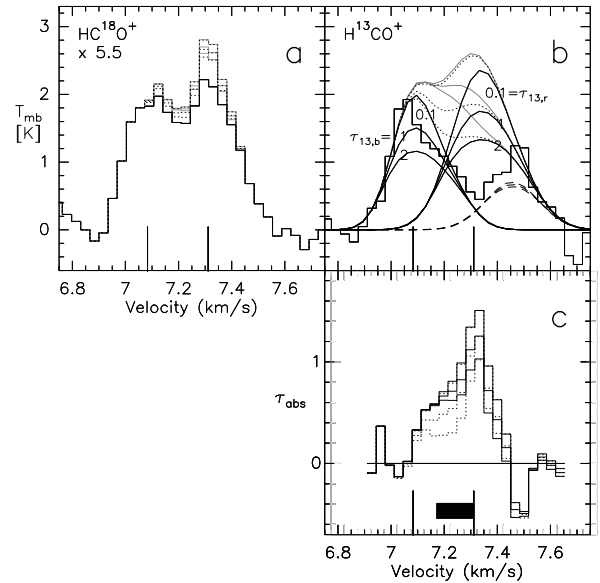}}}
\caption{
Observed L\,1544 spectra (dark histograms) of {\bf a)} HC$^{18}$O$^{+}$(1~--~0) multiplied by 5.5, and of {\bf b)} H$^{13}$CO$^{+}$(1~--~0); after Caselli et al. (2002). 
Vertical bars mark the centers of the ``blue'' and ``red'' velocity
components of HC$^{18}$O$^{+}$(1~--~0) and of numerous other transitions.
Superimposed on b) are synthetic H$^{13}$CO$^{+}$(1~--~0) spectra as extrapolated from each of
these two HC$^{18}$O$^{+}$(1~--~0) components alone (smooth dark lines) and from their
combinations (light colors, see text); assumed H$^{13}$CO$^+$ optical depths are indicated.
Dashed lines give the ``{\it red}'' component's {\it secondary}
contribution alone, without which the measured profile could not be fitted easily (see text).
{\bf c)} Optical depth of absorbing foreground material.
The fat horizontal bar delimits the interval of bulk velocity 
of the absorbing gas.
Light histograms in a) show HC$^{18}$O$^{+}$(1~--~0) profiles corrected for this presumed foreground absorption.
\label{fig6}}
\end{figure}

In the latter, a deep trough is seen that might stem from foreground absorption, while the dip in the
former should rather reflect the separation of the two distinct velocity components mentioned,
because this dip is seen in HC$^{17}$O$^{+}$ as well.

Assuming the two isotopomers' excitation temperatures to be equal and, as also the abundance ratio,
to be constant along the line of sight, we could calculate
the H$^{13}$CO$^{+}$(1~--~0) doublet emission profile that should correspond to the HC$^{18}$O$^{+}$ spectrum,
provided $\tau_{18}$ were known.
Since the optical depths $\tau_{18}$
of both the blue and red velocity components
of HC$^{18}$O$^+$ are a priori unknown, we choose a series of trial values for either
and multiply them by an
abundance ratio $^{13}$C:$^{18}$O like, e.g., 7.3 to obtain $\tau_{13,b}$ and $\tau_{13,r}$.

With these we compute blue and red H$^{13}$CO$^+$(1~--~0) profiles 
like the dark smooth lines of Fig.~6b which, from top to bottom,
correspond to H$^{{\bf 13}}$CO$^+$(1~--~0) optical depths
of 0.1, 1 and 2 as indicated in the figure.
These are {\it total} optical depths, i.e. the primary and secondary hyperfine component taken together.
Three quarters of these values, or 5.5$\times\tau _{18}$, would be the
optical depths of the primary hyperfine component. Combining the blue synthetic
spectrum for optical depth 0.1 (because only low optical depths
satisfy the observation) with the red one for $\tau_{13,r}$ either 0.1, 1 or 2
then gives the total synthetic H$^{13}$CO$^{+}$(1~--~0) profiles (the grey curves); full lines
for the case  that blue and red components are spatially separated
in the beam, dotted lines for blue beyond (and thus partially
absorbed by) red.

Correspondence between the published spectrum and the synthesised ones is
quite close outside the absorbed central velocity interval. In particular,
the high-velocity flank is very well described when hyperfine structure is
taken account of. There is now no need for the H$^{13}$CO$^+$ lines being {\it intrinsically} broader than those of other species, or for
particularly large optical thickness, or an exceptional increase in velocity separation between the
red and the blue components. The cause is the red secondary hf component,
depicted for itself as the dashed curves in Fig.~6b. This varies little with
the particular choice of $\tau_{18,r}$ as optical depth effects would
come into play here only at inappropriately large values of $\tau_{18,r}$.
Any value that makes $\tau_{13,r}$ not much larger than 2 seems compatible
with the data. On the other hand, $\tau_{13,{\bf b}}$ can hardly exceed unity; that
component is likely optically thin.

We may imagine the H$^{13}$CO$^+$ absorption trough to be caused by a sheet of low-excitation
gas placed in front of the emitting volume.
Division of the observed by one of the calculated profiles then leads to
that sheet's optical depth as a function of velocity, at least if the sheet's own emission is negligibly small. This velocity dependence
becomes much clearer after the secondary's contribution to absorption
has been subtracted out. Fig.~6c shows $\tau_{abs}$ (the optical depth
of the foreground {\it primary} alone). $5.5^{-1}$ of it should be the 
absorbing depth of foreground HC$^{18}$O$^{+}$(1~--~0) if the above assumptions continue to hold in the absorbing medium, too. Then,  correction of the observed HC$^{18}$O$^{+}$(1~--~0) profile of Fig.~6a
for this selfabsorption results in the grey step profiles. These in turn entail a slightly different set of synthetic H$^{13}$CO$^{+}$(1~--~0)
profiles; changes are only minute, however.

Quite independently of the value we choose for $\tau_{18,r}$ and of the mode of combining
blue and red in the beam, $\tau_{abs}$ seems composed of two distinct contributions,
a very narrow one of width 0.12~km~s$^{-1}$, i.e. the thermal width of 9~K
gas, and a weaker, broader one. The narrow one is centered exactly on the
velocity center of the red component of the total line profile, the broader one declines smoothly from
there toward lower velocities, disappearing before reaching the blue component's
velocity center.
(Note that with ``velocity center'' and ``linewidth'' we here mean that of the primary
alone; both are identical with those of HC$^{18}$O$^{+}$ but for a 0.004~km~s$^{-1}$
increase in the width).
Since the absorbing linewidth must be at least thermal, the macroscopic gas velocities
{\it excluding thermal} at which foreground absorption takes place are limited to 
the dark bar region in Fig.~6c, between approximately the assumed v$_{LSR}$ of L\,1544 (the mean between
the blue and red line centers) and the red line center itself; nothing beyond on either side.
The blue component seems to be without any absorbing cover of its own, in our direction.

The linewidth of $\approx$ 0.24~km~s$^{-1}$ that both isotopomers show in their red emission component is considerably larger than
the 0.12~km~s$^{-1}$ of the contributor to absorption which is placed exactly at the
velocity center of that red component, velocity-wise. The centrality of this position makes a velocity gradient
a somewhat unlikely cause for the large width of 0.24~km~s$^{-1}$; rather, it seems to speak for
some kind of turbulent origin for this width.

\subsection{Other examples}

Takakuwa, Mikami and Saito (1998)
have presented an interesting comparison between H$^{13}$CO$^{+}$ and CH$_{3}$OH emission from
prestellar dense cores in TMC-1C. They argue convincingly that the two species trace different
stages in the evolution of collapsing clumps since methanol is abundant at early stages
while HCO$^{+}$, being made from CO, accumulates only after CO has been
amply produced.
Consequently they expect somewhat different physical properties, on average, between
the cores displaying predominantly CH$_{3}$OH and those with H$^{13}$CO$^{+}$ being prevalent. Instead, they observe {\it no}
difference in size, linewidth or virial mass between the two groups in spite of their clear chemical
differences.
In particular, mean linewidths are found to be 0.33$\pm$0.11~km~s$^{-1}$ for the ion versus 0.31$\pm$.09
for methanol.

These are narrow widths, hence the H$^{13}$CO$^{+}$ value should be discussed with caution. We have
computed the reduction in true width to be expected when the hyperfine structure of H$^{13}$CO$^{+}$ is fully taken into account, binning our channels
to their velocity resolution of 0.13~km~s$^{-1}$. The result depends on optical
depth. For very low values of $\tau_{13}$, the average width reduction is 0.055~km~s$^{-1}$,
marginally compatible with both chemical groups being of equal widths.
For $\tau_{13}$ values of unity or above, the average width
reduction would be 0.10~km~s$^{-1}$ or roughly one third their value that was obtained without regard to hyperfine structure,
and this would introduce a clear kinematical distinction between early and late clumps.
Also, their virial masses would then have to be reduced correspondingly, halving the mean value
to 1.2 $M_{\odot}$ for the H$^{13}$CO$^{+}$ clumps, just about equal to their mean LTE mass of 1.4 $M_{\odot}$.
Since the damping of turbulence during core evolution poses interesting questions, 
{\it high-resolution}, high signal-to-noise observations of H$^{13}$CO$^{+}$ would seem very desirable.

Another topic where precise interpretation of H$^{13}$CO$^+$(1~--~0) linewidths might become useful concerns
the determination of magnetic fields in dark clouds by comparing the widths of ions and
neutrals as suggested by Houde et al. (2000). Results obtained from the lowest rotational
transitions of H$^{13}$CO$^+$ and H$^{13}$CN (Lai et al. 2003) indicate width differences
in DR21(OH) to be on the order of the ion's hf splitting, hence might be altered (and enhanced)
when this splitting is taken into account.

A related effort by Greaves \& Holland (1999) of comparing H$^{13}$CO$^+$(4~--~3) to $^{34}$CS(7~--~6) velocities, instead of linewidths, seems inconclusive because
their use of a rest frequency more uncertain than ours (Table 3) for the former species results in velocity errors larger than
the velocity differences they measure at various positions.
Indeed, their correlation between this difference and the alignment of grains as derived from
continuum polarization all but vanishes once the new H$^{13}$CO$^+$ rest frequency is employed.

\section{Observation of hyperfine structure in $^{13}$CO(1~--~0)}

Having realised the analytical potential of the hyperfine splitting in H$^{13}$CO$^{+}$ we started to search for
a similar situation in $^{13}$CO.
The lowest rotational transition of this molecule had recently been calculated (see CDMS 2001, based on Klapper et al. 2000) and observed in the laboratory (Cazzoli et al. 2002a) to be splitting into two components,
of frequency difference about 46 kHz (see Table 3), or
also 0.13~km~s$^{-1}$ in velocity space. The line strength ratio is 2:1 in favor of the higher
frequency component. Since $^{13}$CO is a ubiquitous molecule in interstellar
space it would be tempting to apply an analysis comparable to the preceeding one
to the huge amount of available $^{13}$CO data. However, linewidths are usually large
enough to drown any hyperfine effects; consequently, no such investigation seems to
have been attempted to date.

The dark cloud L\,1512 has, as mentioned, zones of exceptionally narrow lines
in numerous species. 
Here, even $^{13}$CO widths can be as low as 0.35~km~s$^{-1}$, sufficiently narrow to warrant
attention to the hyperfine splitting if comparison with other lines is to be carried out. We
therefore searched for hfs effects in the most dominant of these zones. 
We used the extensive IKP CO survey of this object (Falgarone et al. 1998)
which covers a
300$''\times 400''$ field in the relevant transitions.
Nearly the whole
IKP field emits strong (T$_{mb}$ $\approx$ 6~K) $^{13}$CO(1~--~0) radiation of relatively
smooth spatial structure. C$^{18}$O, on the other hand, is more concentrated in a small number of 
clumps, the most prominent of which coincides with the H$^{13}$CO$^{+}$ clump discussed
above.
We chose a 100$''$$\times$100$''$ field centered on this clump to investigate hfs effects in
 $^{13}$CO.
To improve the signal-to-noise ratio we coadded four nearest neighbors in the IKP data set
for all of the following.
For the purpose of this hfs study we make the simplest of possible assumptions, namely that excitation temperature and
beam filling factor
do not vary {\it along} each line of sight.

\subsection{The line profiles}

\begin{figure}[t]
\centering
\resizebox{13.20cm}{!}{\rotatebox[origin=br]{-90}{\includegraphics{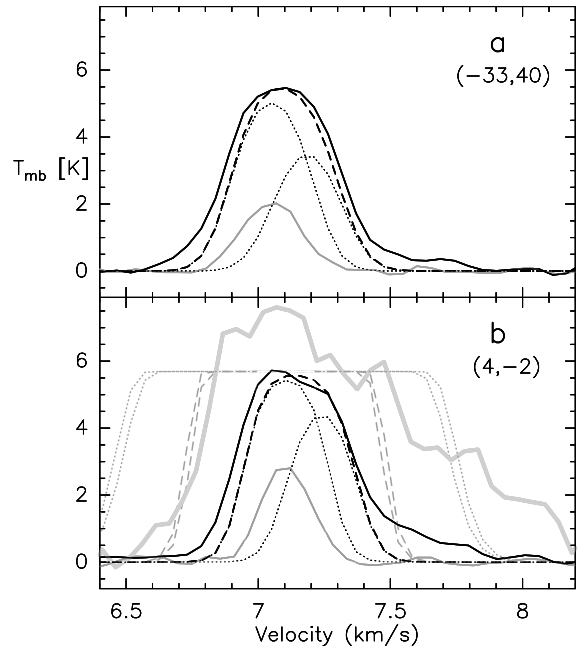}}}
\caption{
L\,1512 spectra of $^{13}$CO(1~--~0) (full black curves) and of C$^{18}$O(1~--~0) (full grey), observed at the offsets indicated (in arcsec).
Dashed black curves show the $^{13}$CO doublet profiles that
would be extrapolated from the C$^{18}$O profiles as described in the text; black dotted lines mark
the individual emissions from their two hyperfine components.
Panel {\bf b)} also displays the $^{12}$CO(1~--~0) emission (heavy light-grey);
the flat-topped dashed grey curves give extrapolations (under simplifying assumptions) from hfs-corrected $^{13}$CO when a $^{12}$CO:$^{13}$CO abundance
ratio of 40 or 70 is applied.
Wide grey-dotted profiles: Same when no hfs correction
is taken into account.
\label{fig7}}
\end{figure}

Quite generally, the $^{13}$CO(1~--~0) and C$^{18}$O(1~-~0) IKP spectra from any one position in this clump do
not, at first sight, seem to have much in common. Assuming singlets for both lines, no reasonable abundance ratio would explain
their large linewidth differences and relative v$_{LSR}$ offsets.
When hyperfine structure is properly taken into account, however,
a close relationship becomes apparent. 
This is demonstrated by the typical line profiles shown in Fig.~7;
a: from near the clump's top edge, b: from close to its center.
Assuming T$_{ex}$ to be equal for both species in such a high density region, and positing a value for {\it A}, the $^{13}$CO to C$^{18}$O abundance ratio,
one could calculate the doublet profiles of $^{13}$CO(1~--~0) that would correspond exactly to the
observed C$^{18}$O profiles, if only $\tau_{18}$, the optical depths of the latter, were known.
By varying $\tau_{18}$ and comparing the $^{13}$CO calculation to observation one obtains an estimate of the actual magnitude of $\tau_{18}$.
Such resulting calculated $^{13}$CO(1~--0) profiles of maximal 18--13 correspondence are entered as
the black dashed lines in Figs. 7a and b.
Although these synthetic $^{13}$CO 
profiles are still marginally narrower than the observed ones, the correspondence is tempting and leaves little room for an all too different emission regime for the two lines.
%(Strictly speaking, if the C$^{18}$O profile consisted of a turbulent superposition of several
% components, of thermal width $\approx$ 0.13~km~s$^{-1}$ each, each of them
%should be treated individually and then combined. Hyperfine splitting, thermal width,
%necessary velocity width of the turbulence spectrum, and optical depth would all cooperate,
%however, to produce practically the same $^{13}$CO profile, of same width, same
%amplitude. This in turn means that turbulence must be studied by other, wider lines).

That the consideration of hyperfine structure may render comparison of $^{13}$CO profiles with other isotopomers more fruitful is underlined in
Fig.~7b where $^{12}$CO(1~--~0) is shown (fat light-grey) as well. Were one to take
the {\it intrinsic} (individual hfs component) linewidth of the best-fit synthetic $^{13}$CO profile and
its best-fit optical depth, and multiplied the latter by a $^{12}$CO:$^{13}$CO abundance
ratio factor of $\approx$ 40 or 70, one would obtain one of the grey dashed profiles as an
expectation for $^{12}$CO(1~--~0), compatible on the low-velocity side with the
actual observation. Were one to disregard hfs and take the full observed width of $^{13}$CO as indication of a high value of $\tau_{13}$ instead,
much wider $^{12}$CO profiles like the dotted ones would result because of the correspondingly higher values of $\tau_{12}$.

Note that at the position near the clump's edge (Fig.~7a), $^{13}$CO observation and
construction are well centered on each other while in the clump's more central region there
is a clear shift between the two.
One possible explanation could be absorption by ``wing'' material that is seen
here in emission above 7.4~km~s$^{-1}$ in both the $^{12}$CO and $^{13}$CO spectra.
Such ``wings'' on $^{13}$CO, present at numerous positions in the core, are a major handicap for
determining $\tau_{13}$, the total peak optical depth of $^{13}$CO, via hfs.
%Depending on the v$_{LSR}$ value at which profile fitting is to be cut off to avoid wing dominance, results will be somewhat different; after all,
%linewidths of 0.35~km~s$^{-1}$ are large relative to a hyperfine splitting of 0.13~km~s$^{-1}$.
Therefore, the following figures are meant to demonstrate trends and possibilities of
analysis rather than to produce hard numbers. We have verified that the {\it trends} depend
little on where wing cutoffs are placed.

\subsection{The potential of hyperfine analysis of $^{13}$CO}

\begin{figure*}[t]
\centering
\resizebox{18.0cm}{!}{\rotatebox[origin=br]{-90}{\includegraphics{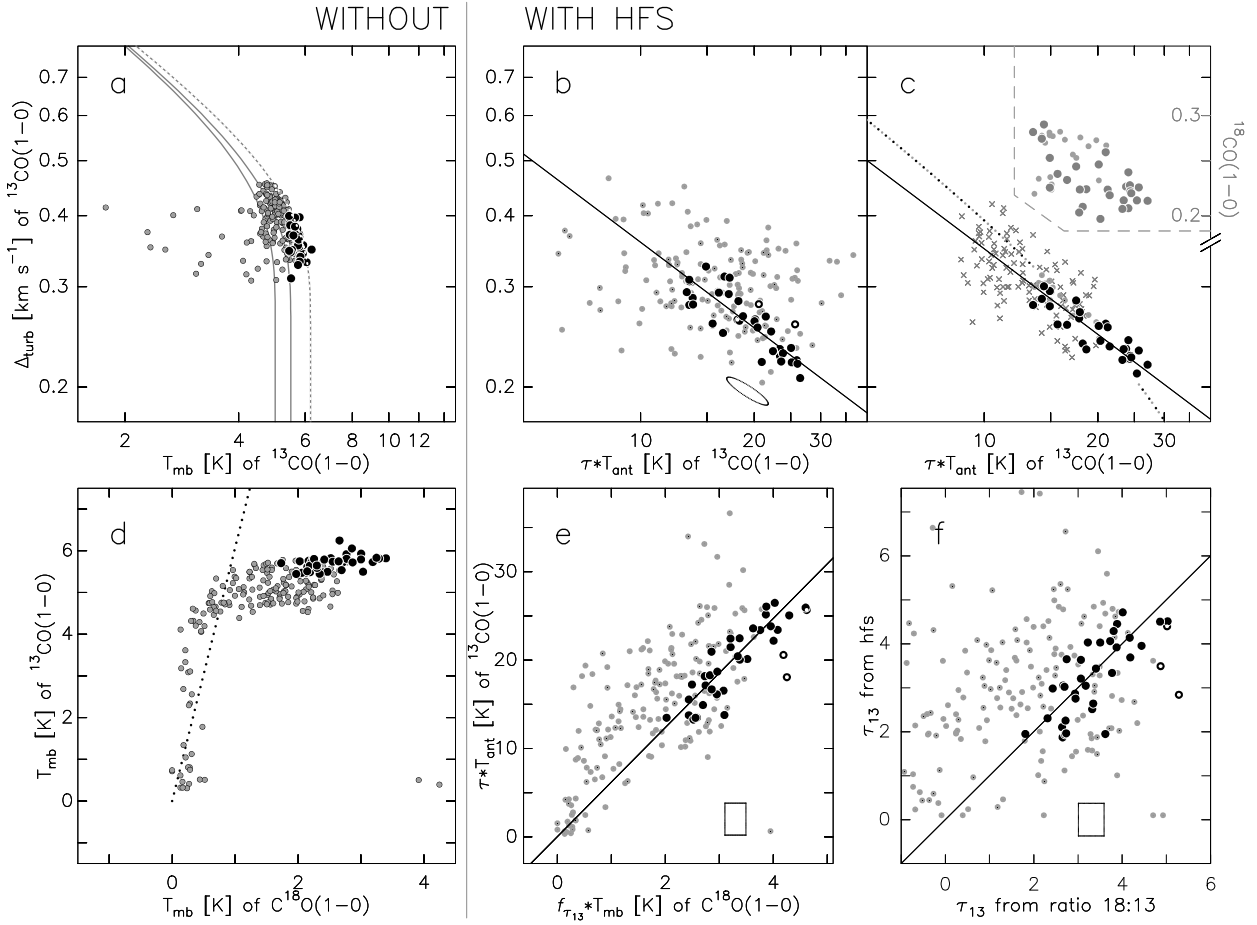}}}
\caption{
Properties of $^{13}$CO(1~--~0) and C$^{18}$O(1~--~0) in the core field (black dots)
and the full IKP field (grey dots) of L\,1512. {\bf a-c)} $^{13}$CO analysed for itself,
{\bf d-f)} in conjunction with C$^{18}$O. Left column: results obtained without considering
hyperfine splitting, right columns: taking hfs into account. Error sizes for the black dots are
given in panels b, e and f by large open symbols.
$\Delta v_{turb}$ is linewidth after substraction of a thermal component
of 10~K.
%T$_{ant}$ as defined in the text.
%
{\bf a)} Smooth curves represent limits derived from the linewidth-optical depth relation.
{\bf b)} Decrease of linewidth with increasing optical depth (note that
T$_{ant}$ hardly varies over the core field). The straight line is the square root
relation between the two quantities.
{\bf c)} Same as b) but with additional consideration of
 C$^{18}$O data.
{\bf d)} Optically thin positions would have to fall along the dotted line {\it if} the $^{13}$CO:$^{18}$CO abundance ratio
were constant at 6.1 over the field.
{\bf e)} Close correspondence between $^{13}$CO and C$^{18}$O emission {\it in the core
region} becomes evident after C$^{18}$O intensity is multiplied by the optical depth factor derived from analysing $^{13}$CO.
{\bf f)} Comparison of two independent methods of determining $^{13}$CO optical
depth (hyperfine analysis vs. amplitude ratio). Note the underabundance of C$^{18}$O outside
the core region, evident in both {\bf e)} and {\bf f)}.
\label{fig8}}
\end{figure*} 

Clearly the extra information contained in $^{13}$CO hyperfine structure provides insights into the physics of L\,1512 that deepen
those already published by Falgarone et al. (1998) and Heithausen et al. (1999).
With $\tau_{13}$ known the physical quantity T$_{ex}$ resp. T$_{ant}$ can now be obtained from T$_{mb}$
according to
T$_{mb}$ = $f_{bf}\cdot $T$_{ant}\cdot $(1-$e^{-\tau})$.
True linewidths of individual components of the doublet can be determined, sometimes giving values considerably smaller than what interpretation in terms of a
single line might suggest.
Such singlet interpretation is bound to also introduce errors in v$_{LSR}$, of up to some 0.04~km~s$^{-1}$ depending on optical depth which influences
the center of gravity of the doublet (cf. Fig. 5).
Finally, the (hidden) correspondence between the seemingly uncorrelated $^{13}$CO and C$^{18}$O lines puts comparisons between
the two isotopomers on a much firmer basis.

We will not reanalyse the IKP data here but only sketch the usefulness of hfs analysis for the L\,1512 core when
its material becomes opaque in the $^{13}$CO~(1~--~0) doublet.
Non-hfs results of
Falgarone et al. (1998) and Heithausen et al. (1999)
are redrawn in Figs. 8a and 8d, distinction now being made between core positions (black dots)
and the whole remaining IKP field (grey)
excepting points of too weak a signal; hollow black points are to be excluded for good reason, e.g. superposition
of two velocities.
Since one-Gauss fits are poor approximations to overlapping doublet profiles, our Figs. 8a and 8d
employ peak intensities and second velocity
moments instead of the corresponding one-Gauss quantities.
(Note that in Figs.~8a-c linewidth is shown as
$\Delta v_{turb}$, i.e. {\it after subtracting out} a
10~K thermal contribution).

Immediately obvious in both figures is the apparent lack of any correlation for the black dots.
This is because all over the core, T$_{ant}$(13) lies in the narrow range
between 5.5 and 6.2~K (which corresponds to T$_{ex}$ between 8.8 and 9.5~K),
and $\tau_{13}$ exceeds 2,
making T$_{mb}$ more or less a surface quantity that is not meaningfully compared to a bulk property
like linewidth.
The inverse linewidth-peak intensity correlation extensively discussed by Falgarone et al.(1998)
must not, therefore, be applied to opaque condensations like the core of L\,1512.
What then could take its place?
The straightforward extension of T$_{mb}$ into the regime of larger optical depths is $\tau \cdot$T$_{ant}$,
a quantity accessible to observation via hfs analysis. 
Substituting this quantity for T$_{mb}$ in Figs. 8a and 8d, and defining
the ordinate $\Delta v_{turb}$ to now mean the {\it intrinsic} linewidth of the {\it individual} hfs
components,
clear correlations then appear as seen in Figs. 8b and 8e.
In particular, since T$_{ant}$(13) has turned out to be very nearly
constant over the core, a relation close to $\Delta v_{turb} \propto \tau_{13}^{-1/2}$ (the straight line in Fig. 8b) emerges.
Also, the non-correlation of the T$_{mb}$ values of the two isotopomers in Fig. 8d transforms into the
linear correlation of Fig. 8e for the core points if a constant value for {\it A} (here: 6.1), hence for the unknown quantity
$\tau_{18}$ (=$\tau_{13} \cdot {\it A}^{-1}$) is assumed.
The factor {\it f$_{\tau_{13}}$} is $A^{-1} \cdot \tau_{13} \cdot (1-e^{-A^{-1} \cdot \tau_{13}})$.
One thus finds T$_{ant}$ of the two species to be within ten percent of each other {\it all over the
core region} while outside,
either $\tau_{18}$ or T$_{ant}$(18) seems to be wanting
with respect to $\tau_{13}$ or T$_{ant}(13)$. 
To the extent that in this high density region excitation temperatures at any given position are the same for the two species,
this indicates that the assumption of a constant $\tau_{18}:\tau_{13}$ ratio is not valid for the IKP field {\it at large};
the abundance ratio there is different from the core's.

How reliable are the $\tau_{13}$ values derived from hfs? Under the assumptions of equal T$_{ant}$ values
for both species (as expected for densities above 10$^4$ cm$^{-3}$ where CO is thermalised) and of a constant abundance (hence $\tau$) ratio,
$\tau_{13}$ can also be derived by the completely independent method of amplitude comparison between the two isotopomers. Results from
these two unrelated methods are juxtaposed in Fig. 8f
where the clustering of the {\it core} points along the equality line shows the effectiveness of the hfs method;
at the same time
the larger IKP field is seen to obviously
be quite underabundant in C$^{18}$O.
This underabundance, which is likely caused by preferential UV destruction of C$^{18}$O in the more
transparent zones exterior to the core, must be the reason why
the positions of low optical depths in Fig. 8d do not accumulate along a straight line like the dotted one (which represents
an abundance ratio {\it A} of 6.1) as they should do if this ratio
were constant across the IKP field.
%Fig.~8f demonstrates that, inspite of the relatively large overall width of the $^{13}$CO profiles
%and concomitant difficulty in determining $\tau_{13}$ well,
%meaningful results can be obtained by hfs fitting alone. In that figure, $\tau_{13}$ as obtained from hfs analysis is compared to
%the values derived from
%the classical method
%It should be noted that the amplitude method {\it alone} works well {\it only} after hfs knowledge has determined which parts
%of the profiles are to be compared with which, and {\it if} the abundance ratio can
%be safely assumed known, a condition not fulfilled over much of the IKP region.
%In conjunction with hfs, however, abundance variations may in principle be estimated everywhere as long as linewidths permit.

%This result has to be taken with caution as the uncertainty of $\Delta v$
%and $\tau_{13}$ (the ellipsoid in Fig 8b) extends just along the correlation line:
%a profile can be fitted somewhat interchangeably by narrow width and high optical
%depth or else by a larger width at lower depth. The
%extent of the error ellipsoid is seen here to lie well below the spread of the data points, however.

\subsection{Relation between linewidth and optical depth}

Since for the core positions both methods of $\tau_{13}$ determination give similar values, they can be
combined to reduce the noise in the final result. The improvement over Fig. 8b is shown in Fig. 8c, dark dots.
With T$_{ant}$(13) practically constant all over the core, the $\Delta v-\tau$ relation (straight line) for this region then reads
$\Delta v_{turb} = const \cdot \tau_{13}^{-1/2}$ with $const = 0.46$~km~s$^{-1}$.
(Note that if $\tau \cdot T_{ant}$ were
a power law function of {\it total} $\Delta v$ rather than of {\it turbulent} $\Delta v$
alone, the black points would accumulate along the dotted curve instead).
Total $^{13}$CO column density $N_{^{13}CO}$ in LTE is
$N_{^{13}CO} = 2.56\times10^{14}\cdot f_{bf}\cdot\tau_{13}\cdot \Delta v\cdot T_{ex}\cdot (1-e^{-T_{o}/T_{ex}})^{-1}$
cm$^{-2}$,
with $\tau_{13}$ the sum of the peak optical depths of the two hfs components and
$\Delta v$ the {\it intrinsic} width ({\it in}cluding thermal) of either of them, measured in km~s$^{-1}$.
At T$_{ex}$ around 9~K this means $N_{^{13}CO}$ around $5.2\times10^{15}\cdot\tau_{13}\cdot \Delta v$ cm$^{-2}$
if $f_{bf}$ is equal to one.
One thus arrives at the approximate further relation $\Delta v_{turb} = 1.13\times10^{15}\cdot N_{^{13}CO}^{-1}$~km~s$^{-1}$.
IKP data on C$^{{\bf 18}}$O (inset) alone would be insufficient for determining such relations.
(Note: for the inset the ordinate is linewidth of C$^{18}$O {\it without }
substraction of a thermal component).

For positions of weaker $^{13}$CO(1~--~0) lines, hfs analysis produces very uncertain values of $\tau_{13}$.
One can make use of the near constancy of T$_{ex}$ over the whole field
by postulating a value of T$_{ex}$, here around 10~K, and then estimating $\tau_{13}$ from
the measured value of T$_{mb}$(13). In Fig.~8c all points obtained in this way are entered as crosses as long as their $\tau \cdot$T$_{ant}$ values are below 20~K.
It thus emerges that the $\Delta v-\tau$ relation extends over much of the whole IKP field,
provided the filling factor is unity everywhere.

With this relation
both $\Delta v_{turb}$ and T$_{mb}$(13) are obtained as a function of
$\tau_{13}$ once a value of T$_{ex}$ is given.
The ensuing $\Delta v_{turb}$-T$_{mb}$ connections are shown in Fig. 8a by smooth curves depicting the cases 
T$_{ex}$(13) equal to 8.3~K, 8.8~K, and 9.5~K (dotted), respectively; the observed points
fall closely between these curves in accordance with hfs analysis which has established this quite narrow T$_{ex}$
regime over most of the field. The Falgarone et al.
(1998) relation $\Delta v \propto T_{mb}^{-1}$ is now recognized as the limiting case of low optical
depths.
%Of course, the proportionality factor between $\Delta v$ and $\tau_{13}^{-1/2}$ is not necessarily the same
%everywhere in the {\it exterior} field. Therefore in Fig.~8a the smooth curve for the regions with T$_{ex}$(13) = 10.3~K, for example,
%would be slightly different from the two shown (but still would nearly coincide with them).

\begin{figure}[t]
\centering
\resizebox{21.2cm}{!}{\rotatebox[origin=br]{-90}{\includegraphics{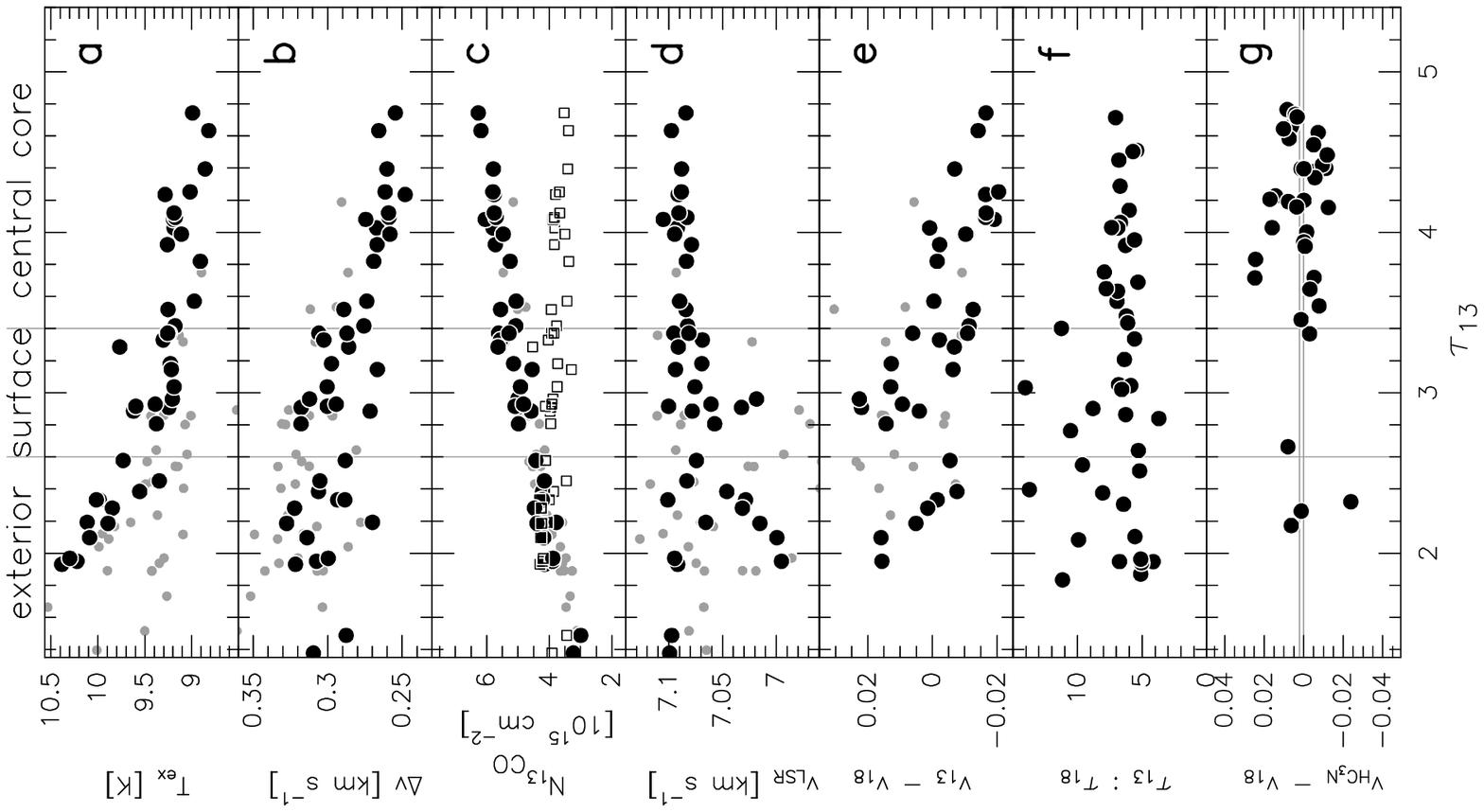}}}
\caption{
Physical properties of the extended core region of L\,1512 (black dots) and of the IKP field at large (grey), as a function of
  $^{13}$CO optical depth.
  The two vertical lines
  refer to the black dots only.
Panels {\bf a-d} are derived from $^{13}$CO alone, {\bf e-f} from $^{13}$CO and C$^{18}$O combined.
  {\bf a)} Excitation temperature, derived from T$_{ant}$ by assuming beam filling factor unity.
{\bf b)} {\it Total} linewidth, i.e. {\it in}cluding
  thermal, of a single hf component.
{\bf c)} Column density N$_{^{13}CO}$, derived under the assumption of beam filling unity. 
  Open squares show this quantity as determined without recourse to hyperfine analysis,
dots the results obtained when employing hfs.
{\bf d)} v$_{LSR}$ of $^{13}$CO(1~--~0), and {\bf e)} relative motion between
  $^{13}$CO and C$^{18}$O,
both in km~s$^{-1}$.
{\bf f)} Quotient of optical depths, indicative of the abundance ratio $^{13}$CO:C$^{18}$O.
{\bf g)} v$_{LSR}$ difference between 
  HC$_{3}$N(3~--~2) 
  and C$^{18}$O(1~--~0)
at selected positions in the core, in km~s$^{-1}$;
upper horizontal line is the mean value of the difference.
\label{fig9}}
\end{figure} 

One simple picture to explain our power law relations would be of a quiescent,
narrow-line core surrounded by a turbulent envelope, both of beam filling factor unity. %In the framework of a spherically
%symmetric model one can show however that, independent of the radial variations of
%density and turbulent width, it is not possible to construct solutions with power law
%exponents as large as found here, over such a wide span of optical depth. A more general
%approach is warranted; the most obvious generalization
%would seem to be small-scale inhomogeneities with filling factors decreasing from
%inside out. 
Goodman et al. (1998) gave a thorough discussion of the transition from coherence to
incoherence in dense cores.
Also of particular relevance here are the recent observations by Kim \& Hong (2002) who
in numerous dark clouds
found nearly {\it no} variation at all of the turbulent velocity dispersion with distance from center.
Note that our results to the contrary, at least in the core of
L\,1512, would be largely washed out if we smeared out the IKP beam to their
angular resolution of 50$''$. Our inside-out increase
of $\Delta v_{turb}$ might be direct indication of the transition from a very quiescent
nucleus to its turbulent envelope. Measurements in other objects would be
desirable.

Finally, the slight width difference
between the synthetic and the observed $^{13}$CO profiles of Figs. 7a and b may also find its natural
explanation in the width-optical depth relation.
It is in the outer, less opaque zones that the widest contributions to the line profiles are
produced. These are also the zones of C$^{18}$O underabundance. Hence relatively less widening
in this isotopomer's lines, thus also in the synthetic $^{13}$CO profiles derived from them, is to be expected.

\subsection{Some physical properties of L\,1512 as derived from hyperfine analysis}

Fig.~9 summarises some physical trends, as recognised by hyperfine structure analysis, of the core {\it plus}
its immediate ($\approx 25 ''$ wide) surroundings. The central core here
is defined to comprise all positions with values of $\tau_{13}$ larger than 3.4, its immediate surface
those between 2.6 and 3.4, and the exterior the values below. The top four panels result from $^{13}$CO hfs analysis alone, the lower ones 
from a combination with C$^{18}$O.

Excitation temperatures (top panel) tend to
decrease from outside in, as do the linewidths just discussed at some length.
T$_{ex}$ here is derived under the assumption of a constant beam filling factor equal to one.
Since this might not hold in the surface region,
the true value of T$_{ex}$ there could be somewhat higher.
Note that the excitation temperature of the main isotopomer, $^{12}$CO, is at
11.0~K all over the core's surface (cp. Fig.~7b).

Column density (Fig.~9c; unity beam filling factor assumed) increases smoothly with $\tau_{13}$.
In contrast, it would seem nearly constant (the open squares in the panel) if the standard method of deriving
this quantity were followed. That method assumes equality of the excitation temperatures of
$^{13}$CO and $^{12}$CO (the latter being measurable directly because of very large optical
depths), and estimates $\tau_{13}$ by comparing the T$_{mb}$ values of the two isotopomers. The
assumption on T$_{ex}$ is clearly not valid here, however, since $^{12}$CO is mainly emitted
from the hotter surface, $^{13}$CO from the core. The difference makes for a factor up to 3
in optical depth. The corresponding difference in column density is somewhat lessened by
the opposing difference in the values of linewidth (and in the T$_{ex}$($^{13}$CO)-dependent
factor in $N_{^{13}CO}$), but even so, a profile interpretation employing hfs analysis arrives
at column densities for the central core that are close to twice the ``standard'' values. This means
similarly higher core mass. On the other hand, hfs analysis results in lower intrinsic
linewidths, hence lower virial masses. Therefore, the net effect of taking account of the
hyperfine splitting here is to make cores like L\,1512 appear considerably more stable than when ignoring the splitting.

Fig. 9d: v$_{LSR}$ of $^{13}$CO is at a
well defined constant value for all positions but a few nearer the edge. However, close inspection shows the v$_{LSR}$ {\it difference} (Fig.~9e)
between the 13 and 18 isotopomers to vary from outside in, albeit very weakly;
since the absolute v$_{LSR}$ values are somewhat uncertain in the IKP data set, it is hard to say which 
species is more redshifted than the other. Since in the outer, more
transparent regions both signals will stem to a larger extent from the same gas than further in, it is reasonable to
suspect the central core (more visible in C$^{18}$O than in $^{13}$CO)
to be {\it red}shifted with respect to the surface. More precise measurements would be needed
to quantify.
Recall that at this level of variation, a sufficiently accurate determination of v$_{LSR}$(13) is only achievable by taking hyperfine structure into account.

The ratio of $^{13}$CO:C$^{18}$O optical depths (Fig.~9f) directly reflects, in the central core,
the abundance ratio of the two species. Higher values outside indicate C$^{18}$O destruction beyond the region
protected by large extinction.

Finally, Fig.~9g compares v$_{LSR}$ of C$^{18}$O(1~--~0) (as obtained from a small data set taken independently from
IKP, of lower signal-to-noise ratio but more precise velocity determination) to that of the HC$_{3}$N(3~--~2) triplet mentioned above, at selected positions in or
near the inner core. HC$_{3}$N optical depth in the core is found to be mostly on the order of 40 percent of that of $^{13}$CO,
therefore in HC$_{3}$N and C$^{{\bf 18}}$O one might see about the same regions with similar weight. In fact, no
systematic velocity difference is recognisable at all between these two species except that their mean values differ by
.0014~km~s$^{-1}$ (the upper horizontal line) if the most precise measured C$^{18}$O frequency (Cazzoli et al. 2003) is employed.
This difference is not larger than the published uncertainties of either line and thus lends support to the accuracy of the
H$^{13}$CO$^+$(1~--~0) frequency that was derived in section 6 via HC$_3$N.

\section{Conclusions}

We wanted to show in this paper that the $^{13}$C nucleus' magnetic moment can provide
a useful additional tool for analysing radiation
from both the H$^{13}$CO$^{+}$ and the $^{13}$CO molecules. These species have not previously
been investigated with consideration of this important detail. This is
because the separation of their hyperfine components, 0.13~km~s$^{-1}$ for both species,
is as narrow as the thermal width of cold interstellar gas and therefore a
rather quiescent cold astronomical target, high spectral resolution, and
good signal-to-noise ratios are all required to observe the splitting. When these
conditions are fulfilled, however, narrowness of separation is hardly a
handicap: while partial overlap of the components does of course
degrade information near the profile's center, the noise in the totality of
the line is also less than if the components were distinctly apart. One
should also note that line strength ratios 3:1 or 2:1, as in the present cases,
are next to ideal for retrieving astronomical quantities like optical depth.
Ratios much closer to one would not allow clear determination of depth by
comparison of amplitudes of the hyperfine components, much larger ones on the other hand would carry the
risk of correlating gas regimes that might in reality be quite distinct, one of
high optical depth, the other transparent.

For narrow-line regions, the determination of velocities profits most directly
from knowing the hyperfine structure underlying a profile.
Therefore we were able to present H$^{13}$CO$^{+}$ rotational transition frequencies
of considerably improved precision by comparing to other, well-defined lines
in L\,1512, as well as to verify astronomically the $^{13}$CO(1~--~0) doublet frequencies
first predicted by theory. This precision in turn allows studies of the internal
motions of very quiescent clouds that may be extremely subsonic but, as was shown above
for the case of L\,1512, may nevertheless indicate the onset of collapse toward the center of
a potentially star-forming core.

The inadequacy of interpreting a close-doublet profile by singlet analysis becomes very
evident in the derivation of linewidths. Here, neglect of hyperfine structure in the profile can
easily entail considerable overestimates which in turn lead to erroneous
values of column density and of virial mass (the latter being proportional to the
square of intrinsic linewidth). This effect is compounded by the frequently necessary neglect
of line broadening caused by nonzero optical depth.
In the case of the two $^{13}$C-containing species discussed here, the direct handle on optical
depth that is provided by the nuclear magnetic moment may permit a more reliable
determination of the intrinsic linewidth and its derivatives. For this reason, we were
able to establish a width-depth or width-column density relation for the
core of L\,1512 that may eventually make the transition from a completely quiescent
central core to a turbulent envelope accessible to observations. 
This more direct method of obtaining optical depths, rather than by the standard method
for $^{13}$CO, allowed us to recognise the L\,1512 core as considerably more bound than
suspected since core mass has to be revised upwards and virial mass downwards upon
considering line multiplicity correctly.
Likewise we
tried to argue that measurements in the dark core of L\,1544 which seem to set H$^{13}$CO$^{+}$ quite
apart from numerous other molecular species may in reality fall nicely in line when
hyperfine splitting is properly considered.

A reestimate of H$^{13}$CO$^+$ linewidths also suggests that chemical evolution of
collapsing cores over time should indeed be observable through the comparison of these linewidths
with those of other molecules. Further, recent attempts to derive magnetic field
strengths in molecular clouds by comparing ion linewidths with those of neutrals
(H$^{13}$CO$^+$ vs. H$^{13}$CN) should profit from taking the doublet nature of the ion into consideration.
For all these reasons it is hoped that the hyperfine structure in the
profiles of the species discussed in this paper will motivate further high-resolution
studies of cold dark clouds.

\begin{acknowledgements}
We are very grateful to Michael Grewing for providing director's time, and
to Clemens Thum for arranging the observations, at the 30m telescope of IRAM.
Malcolm Walmsley, Floris van der Tak, and the anonymous referee have contributed very useful comments on the original manuscript.
We also want to thank Otmar Lochner for his help with determining the
frequency accuracy of the 100m telescope system of the Max-Planck-Institut f\"{u}r Radioastronomie.

The work in K\"oln was supported by the Deutsche Forschungsgemeinschaft via grant SFB494 and
by special funding from the Science Ministry of the Land Nordrhein-Westfalen.
\end{acknowledgements}


\begin{thebibliography}{}
\bibitem[1]{1}
Bogey M., Demuynck C., Destombes, J. L. 1981, Mol. Phys., 43, 1043
\bibitem[i]{i}
%Haese N. N., Woods R. C. 1979, Chem. Phys. Lett., 61, 396
Botschwina P., Horn M., Fl\"{u}gge J., Seeger S. 1993, Faraday Trans., 89, 2219
\bibitem[2]{2}
Caselli P., Myers P. C., Thaddeus P. 1995, ApJ, 455, L77
\bibitem[2a]{2a}
Caselli P., Walmsley C. M., Zucconi A., et al. 2002, ApJ, 565, 331
\bibitem[3]{3}
Cazzoli G., Puzzarini C., Lapinov A. V. 2003, ApJ, 592, L95
\bibitem[4]{4}
Cazzoli G., Dore L., Cludi L. et al. 2002a, J. Mol. Spectrosc., 215, 160
\bibitem[d]{d}
Cazzoli G., Dore L., Puzzarini C., Beninati S. 2002b, Phys. Chem. Chem. Phys., 4, 3575
\bibitem[5]{5}
CDMS (2001): Cologne Database for Molecular Spectroscopy:
M\"{u}ller H. S. P., Thorwirth S., Roth D. A., Winnewisser G. 2001, Astron. Astrophys. 370, L49
(http://www.ph1.uni-koeln.de/vorhersagen/index.html)
\bibitem[6]{6}
Ciolek G. E., Basu S. 2000, ApJ, 529, 925
\bibitem[7]{7}
Coxon J. A., Hajigeorgiou P. G. 1992 Can. J. Phys., 70, 40
\bibitem[8]{8}
Creswell R. A., Winnewisser G., Gerry M. C. L. 1977 J. Mol. Spectrosc., 65, 420 
\bibitem[9]{9}
Dore, L., Puzzarini, C., Cazzoli, G. 2001, Can. J. Phys., 79, 359
\bibitem[a]{a}
Ebenstein, W. L., Muenter, J. S. 1984, J. Chem. Phys., 80, 3989
\bibitem[b]{b}
Falgarone E., Panis J. F., Heithausen A., Perault M., Stutzki J., Puget J. L.,
Bensch F. 1998, Astron. Astrophys. 331, 669
\bibitem[c]{c}
Fiebig D. 1990, PhD Thesis Universit\"{a}t Bonn
\bibitem[e]{e}
Garvey, R. M., De Lucia, F. C. 1974, J. Mol. Spectrosc., 50, 38
\bibitem[f]{f}
Goodman A. A., Barranco J. A., Wilner D. J., Heyer M. H. 1998, ApJ, 504, 223
%//Gregersen E. M., Evans II N. J. 2000, ApJ, 538, 260
\bibitem[g]{g}
Greaves J. S., Holland W. S. 1999, MNRAS, 302, L45
\bibitem[ga]{ga}
Gregersen E. M., Evans II N. J. 2001, ApJ, 553, 1042
\bibitem[h]{h}
Gu\'{e}lin M., Langer W. D., Wilson R. W. 1982, Astron. Astrophys. 107, 107
\bibitem[j]{j}
Heithausen A., Stutzki J., Bensch F., Falgarone E., Panis J. F. 1999, Rev. Mod. Astron., 12, 201
\bibitem[k]{k}
Helgaker T., Jensen H. J. Aa., J{\o}rgensen P., et al.
{\it Dalton release 1.0 (1997), an electronic structure program}
\bibitem[l]{l}
Hirota T., Ikeda M., Yamamoto S. 2003, ApJ, 594, 859
\bibitem[m]{m}
Houde M., Bastien P., Peng P., Phillips T. G., Yoshida H. 2000, ApJ, 536, 857
\bibitem[mm]{mm}
Irvine W. M., Goldsmith P. F., Hjalmarson A. 1987, in {\it Interstellar Processes},
eds. D. Hollenbach and H. A. Thronson (Dordrecht: D. Reidel), 561
\bibitem[n]{n}
Jensen H. J. Aa., J{\o}rgensen P., Agren H., Olsen J.
1988a J. Chem. Phys., 88, 3834
\bibitem[o]{o}
Jensen H. J. Aa., J{\o}rgensen P., Agren H., Olsen J.
1988b J. Chem. Phys., 89, 5354
\bibitem[p]{p}
Kim H. G., Hong S. S. 2002, ApJ, 567, 376
\bibitem[q]{q}
Klapper, G., Lewen, F., Gendriesch, R., Belov, S. P., Winnewisser, G.
2000, J. Mol. Spectrosc., 201, 124
\bibitem[r]{r}
Lai S.-P., Velusamy T., Langer W. D. 2003, preprint
\bibitem[rr]{rr}
Klapper, G., Lewen, F., Gendriesch, R., Belov, S. P., Winnewisser, G.
2001, Z. Naturforsch., 56a, 329
%\bibitem[s]{s}
%Lee C. W., Myers P. C., Tafalla M. 1999, ApJ, 526, 788
\bibitem[t]{t}
Maki A. G., Mellau G. Ch., Klee S., Winnewisser M., Quapp W. 2000, J. Mol. Spectrosc., 202, 67
\bibitem[u]{u}
Muders D. 1995, PhD Thesis Universit\"{a}t Bonn
\bibitem[v]{v}
Myers P. C., Benson P. J. 1983, ApJ, 266, 309
\bibitem[w]{w}
Pickett H. M., Poynter R. L., Cohen E. A., Delitsky M. L., Pearson J. C., M\"{u}ller H. S. P.
1998, J. Quant. Spectrosc. Radiat. Transfer, 60, 883 
\bibitem[x]{x}
Snyder L. E., Hollis J. M., Lovas F. J., Ulich B. L. 1976, ApJ, 209, 67
\bibitem[y]{y}
Tafalla M., Mardones D., Myers, P. C., Caselli P., Bachiller R.,
Benson P. J. 1998, ApJ, 504, 900
\bibitem[z]{z}
Takakuwa S., Mikami H., Saito M. 1998, ApJ, 501, 723
\bibitem[10]{10}
Thorwirth S., M\"{u}ller H. S. P., Winnewisser G. 2000, J. Mol. Spectrosc., 204, 133
\bibitem[11]{11}
Winnewisser M., Winnewisser B. P., Winnewisser G. 1985, in {\it Molecular Astrophysics, Series C},
eds. G. H. F. Diercksen, W. F. Huebner, and P. W. Langhoff (Dordrecht: D. Reidel), 375
\bibitem[11w]{11w}
Woods R. C., Saykally R. J., Anderson T. G., Dixon T. A., Szanto P. G. 1981, J. Chem. Phys., 75, 4256
\end{thebibliography}
\end{document}